\documentclass[astrosymb,twocolumn]{aastex631}

\newcommand{\Msun}{$\rm M_{\odot}$}
\newcommand{\Mcrit}{${M_{\rm crit}}$}
\newcommand{\JLW}{${J_{\rm LW}}$}
\newcommand{\Htwo}{$\rm {H_{2}}$}
\newcommand{\Mmol}{${M_{\rm H_{2}}}$}
\newcommand{\vbc}{$v_{\rm bc}$}
\newcommand{\sigvbc}{$\sigma_{\rm vbc}$}
\newcommand{\tdelay}{$t_{\rm delay}$}
\usepackage{breqn}
\DeclareMathOperator\erf{erf}

\usepackage{graphicx}
\usepackage{units}
\usepackage{sidecap}
\usepackage{longtable}
\usepackage{amsfonts}
\usepackage{wrapfig}

\begin{document}

\title{A Global Semi-Analytic Model of the First Stars and Galaxies Including Dark Matter Halo Merger Histories}

\author[0000-0001-9062-4615]{Colton R. Feathers}
\email{colton.feathers@rockets.utoledo.edu}
\author[0000-0002-9789-6653]{Mihir Kulkarni}
\email{mihir.kulkarni@utoledo.edu}
\author[0000-0002-8365-0337]{Eli Visbal}
\email{elijah.visbal@utoledo.edu}
\author[0000-0002-1034-7986]{Ryan Hazlett}
\email{ryan.hazlett@rockets.utoledo.edu}
\affiliation{University of Toledo,
Department of Physics and Astronomy and Ritter Astrophysical Research Center,
2801 W. Bancroft Street,
Toledo, Ohio 43606}

\begin{abstract}
We present a new self-consistent semi-analytic model of the first stars and galaxies to explore the high-redshift ($z\geq15$) Population III (PopIII) and metal-enriched star formation histories. Our model includes the detailed merger history of dark matter halos generated with Monte Carlo merger trees. We calibrate the minimum halo mass for PopIII star formation from recent hydrodynamical cosmological simulations that simultaneously include the baryon-dark matter streaming velocity, Lyman-Werner (LW) feedback, and molecular hydrogen self-shielding. We find an overall increase in the resulting star formation rate density (SFRD) compared to calibrations based on previous simulations (e.g., the PopIII SFRD is over an order of magnitude higher at $z=35-15$). We evaluate the effect of the halo-to-halo scatter in this critical mass and find that it increases the PopIII stellar mass density by a factor of ${\sim}1.5$ at $z\geq15$. Additionally, we assess the impact of various semi-analytic/analytic prescriptions for halo assembly and star formation previously adopted in the literature. For example, we find that models assuming smooth halo growth computed via abundance matching predict SFRDs similar to the merger tree model for our fiducial model parameters, but that they may underestimate the PopIII SFRD in cases of strong LW feedback. Finally, we simulate sub-volumes of the Universe with our model both to quantify the reduction in total star formation in numerical simulations due to a lack of density fluctuations on spatial scales larger than the simulation box, and to determine spatial fluctuations in SFRD due to the diversity in halo abundances and merger histories.
\end{abstract}


\section{Introduction} \label{sec:intro}
Theoretical calculations based on the standard model of cosmology predict that the first stars  began illuminating the Universe within the first $\sim$100 Myr that followed the Big Bang. Simulations indicate these stars, categorized as Population III (PopIII) stars, formed from primordial metal-free gas within dark matter ``minihalos'' ($M_{\rm vir} = 10^5 - 10^6$ \Msun), with much higher masses ($M_{*} = 10 - 1000$ \Msun) than metal-enriched stars \citep[for a recent review, see][]{PopIII_review}. Thus, the first PopIII stars likely had stellar lifetimes of only a few Myr \citep{Schaerer02}. Following their short lives, they injected metal-enriched material into their surroundings via supernovae (SNe) winds, resulting in the formation of Population II (PopII) stars \citep[see e.g.,][]{2015MNRAS.452.2822S} and ultimately the first galaxies.

While there are currently no confirmed detections of PopIII stars, a variety of upcoming observations have the potential to constrain their abundance and properties (e.g., the PopIII initial mass function (IMF)). For example, the recently launched \emph{James Webb Space Telescope} (\emph{JWST}) may observe pair-instability SNe \citep{Hartwig18b, Whalen13} from PopIII stars with initial masses of $\sim$140 -- 250 \Msun. If large clusters of PopIII stars form in so-called PopIII galaxies (e.g., due to inefficient mixing of metals), they may be directly observable with \emph{JWST} \citep{2017MNRAS.469.1456V, 2019ApJ...882..178K,2019ApJ...871..206S, 2022MNRAS.512.3030V}. We note that large PopIII starbursts detectable with \emph{JWST} have recently been predicted to occur in alternative dark matter scenarios that suppress small scale structure, such as ``fuzzy dark matter'' \citep[e.g.][]{2022ApJ...941L..18K, FDMex2, FDMex, FDM}. While PopIII stars currently remain elusive, \emph{JWST} has already observed metal-enriched galaxies out to very high redshifts \citep[e.g.][]{Naidu22, Labbe23, Finkelstein23}. The properties of low-mass galaxies may even depend on the characteristics of PopIII stars due to the hierarchical nature of structure formation in $\Lambda$CDM \citep[e.g.,][]{2021MNRAS.508.3226A}.

Another promising route to constrain properties of the first stars and galaxies is to measure their impact on the thermal and ionization properties of the intergalactic medium (IGM). For example, the optical depth due to electron scattering of the cosmic microwave background (CMB) has been used to put upper limits on the efficiency of PopIII star formation \citep[e.g.][]{Haiman03, Shull08, Ahn12, Visbal15b, Miranda17}. Additionally, the cosmological 21cm signal is sensitive to the properties of early star formation. This is true both for global experiments such as  EDGES \citep{2018Natur.555...67B} and LEDA \citep{2018MNRAS.478.4193P}, as well as those aiming to measure 3-dimensional (3D) spatial fluctuations, such as HERA \citep{2017PASP..129d5001D} and SKA \citep{2013ExA....36..235M}. Additional observational probes of PopIII stars include gamma-ray bursts \citep{Bromm07, Burlon16, Kinugawa19}, line-intensity mapping of the 1640 \r{A} He II recombination line \citep{Visbal15a, HeIIb}, and studying extremely metal-poor stars in the local universe that could have been formed from the SNe ejecta of PopIII stars \citep[``stellar archaeology" e.g.,][]{Magg18, Hartwig18, Hartwig15, Frebel15}. 

The measurements described above require accurate theoretical models of early star and galaxy formation to both suggest optimal observing strategies and interpret the results. Previous modelling efforts have taken many forms, from large-scale cosmological hydrodynamic simulations \citep[e.g.][]{Renaissance13, Jaacks18} to analytic calculations for quick predictions of global properties \citep[e.g][]{Haiman06, Wyithe07, Trenti09a, Visbal15b, Furlanetto21}. While hydrodynamical cosmological simulations, such as the Renaissance Simulations \citep{Renaissance13, Renaissance14, Renaissance15, Renaissance18, Renaissance16, Renaissance15b}, include most of the physical processes relevant to early star formation, such as radiative, mechanical, and chemical feedback, they are very numerically expensive (e.g., ${\sim}10^7$ CPU hours per realization in the case of Renaissance). While these simulations follow the physics with high fidelity, they still typically involve free parameters in sub-grid prescriptions (e.g., the minimum metallicity for PopII star formation or the PopIII IMF), and are too costly to explore the parameter space in detail. 

Semi-analytic models of the first stars and galaxies represent an alternate approach that can be used to rapidly survey the uncertain parameter space, while accurately following the hierarchical assembly of dark matter structure \citep[e.g.,][]{Trenti09a, Yung1, Crosby13, Valiante16}. The dark matter ``backbone" can be computed either through cosmological N-body simulations or analytic Monte Carlo (MC) methods based on the Extended Press-Schechter (EPS) formalism \citep{Press74, EPSformalism}. Star formation and astrophysical feedback is then implemented on top of this structure using analytic techniques, allowing for rapid simulation across cosmological volumes and large redshift ranges while still including most of the important physics \citep[see][for examples]{Magg18, Mebane18, Visbal18, Visbal20, Liu20}.

In this paper, we present a new global high-redshift ($z\geq15$) semi-analytic model of the first stars and galaxies. The main new features of this model are that it includes complete dark matter (DM) halo merger histories (from MC methods) and is calibrated to state-of-the-art hydrodynamical cosmological simulations of the critical halo mass for PopIII star formation \citep{Kulkarni21}. We utilize our model to make updated predictions of the high-redshift PopIII and metal-enriched star formation rate densities (SFRDs) and to test the impact of various physical effects/parameters (e.g., the baryon-dark matter streaming velocity). Given that our model incorporates fully detailed DM halo merger history, we are also able to test the effect of approximations that have been adopted previously in literature to estimate the SFRD, such as smooth merger histories determined by halo abundance matching \citep[e.g.][]{Furlanetto17,Mebane18} and simple integrals of the halo mass function \citep[e.g.,][]{Visbal15b, Munoz22, Munoz23}. Finally, we use our model to simulate sub-regions of the Universe with different volumes, accounting for the Poisson fluctuations in the number of halos and their diverse merger histories. This allows us to estimate the impact of merger history on 3D spatial fluctuations of the PopIII and metal-enriched SFRDs relevant to future predictions of the cosmological 21cm signal. It also allows us to show the impact on the SFRDs predicted by numerical simulations of finite box size that effectively ignore density fluctuations on spatial scales larger than the simulation box.

The remainder of this paper is structured as follows. In Section \ref{sec:prevworks}, we contextualize our new model by reviewing previous work in the literature. In Section \ref{sec:methods}, we discuss the details of our fiducial model and the physical processes included. The results of this work are divided into Sections \ref{sec:resultsI}, \ref{sec:resultsII}, and \ref{sec:resultsIII}. Section \ref{sec:resultsI} presents the main results of our fiducial model, and details the impact of various modelling fits and parameters on the global SFRD. Section \ref{sec:resultsII} focuses on the impact of modelling choices for the dark matter halo evolution and the prescription for determining PopII star formation. Here we also check the accuracy of previous assumptions made in the literature regarding the effects of halo assembly on predicted SFRDs. In Section \ref{sec:resultsIII} we simulate finite, physically representative volumes of the Universe using our fiducial model. We compare our fiducial SFRDs with those previously published in the literature in Section \ref{sec:discussion}. Finally, in Section \ref{sec:conclusions} we summarize and discuss our results and explore future research. Unless otherwise stated, all distances are in comoving units. Throughout this work, we use a $\Lambda$CDM cosmology, consistent with \cite{Planck} \ and with the following parameters: $\Omega_{\rm m} = 0.32$, $\Omega_{\rm \Lambda} = 0.68$, $\Omega_{\rm b} = 0.049$, $h = 0.67$.

\section{Previous Works} \label{sec:prevworks}
In this section, we briefly review semi-analytic work in the literature to provide context for our new model. The general approach in semi-analytic modelling of galaxy evolution is to generate dark matter halo merger trees, either with MC methods or cosmological N-body simulations, and then apply analytic prescriptions for star formation and astrophysical feedback processes. This has been widely used to model galaxies at lower redshifts than those explored here \citep[for a review see][]{2015ARA&A..53...51S}. It has also been used to study high-redshift galaxies and reionization, but without including PopIII stars \citep[e.g.,][]{Poole16, Mutch16, Yung2, Yung3, Yung4, Yung5, Yung6}. 

Described in detail in Section \ref{sec:methods}, we have adopted this framework to develop a global model of the first PopIII stars and early galaxies that includes full dark matter halo merger history, incorporates self-consistent LW feedback \citep{Haiman97, Haiman00, Machacek01, Oshea08, Ahn09}, and is calibrated to recent hydrodynamical simulations. Most of the previous merger tree-based semi-analytic modelling of the first stars has been utilized to study the assembly of individual halos, rather than the global quantities which we focus on. This includes the formation of high-redshift supermassive black holes in rare high-mass halos \citep[e.g.,][]{Valiante16, 2021MNRAS.503.5046L, Trinca22} as well as Milky Way-like halos to make predictions for stellar archaeology \citep[e.g.,][]{Magg18, Griffen18, Ishiyama16, deBennassuti17, Graziani17}. We note that there have been previous global merger-tree-based models related to ours. In \cite{2016MNRAS.462.3591M}, the PopIII SFRD was the focus; this is updated here with newly simulated critical halo masses for PopIII star formation \citep{Kulkarni21}. More recently, \cite{Trinca22} utilized EPS merger trees to study black hole and PopIII star formation, but this work focused on lower redshifts than what are explored here. We also note that towards the completion of this work, \cite{Ventura22} and \cite{Trinca23} presented similar global models of the first stars and galaxies. Their respective focuses were on the predicted 21cm signal and UV luminosity functions, whereas we utilize our model to explore the impact of various physical effects and modelling techniques on the PopIII and metal-enriched SFRDs. An additional novel aspect of our work is that we use our model to understand the behavior of various sub-volumes of the Universe as a step towards accurate modelling of the 3D spatial fluctuations of high-redshift star formation.  

Several recent semi-analytic works have included 3D spatial information from N-body simulations (in addition to merger trees) to model the effects of ionization feedback and ``external'' metal-enrichment between neighboring halos \citep[e.g.,][]{Sarmento19, Liu20, Visbal20, Hartwig22}. This is complementary to the work presented here. Although our fiducial model cannot properly include 3D feedback (since MC EPS methods are used to generate the merger trees), it has been found that 3D feedback does not have a significant impact of the SFRDs at $z \gtrsim 15$  \citep{Visbal20}, which is when we have decided to stop our models presented below. With our semi-analytic framework, we are able to rapidly explore much larger volumes than what is possible when N-body simulations are used. 

In addition to these semi-analytic methods, analytic work has been used to model the global abundance of the first stars and galaxies. These include what we refer to as ``smooth accretion models'' where halo growth histories are approximated using abundance matching of the halo mass function \citep{Furlanetto17, Mebane18, Mebane20}. A number of previous works have also estimated the SFRDs of PopIII and metal-enriched stars via analytic integration of the halo mass function \citep[e.g.,][]{Visbal15b, Munoz22, Munoz23}. We use our model below to test the impact that these approximations related to halo assembly have on the global star formation history.

We also note that our new model relates to large-scale semi-numerical models which have been used to predict the high-redshift 21cm signal \citep{21cmFAST, McQuinn12, 2012Natur.487...70V, Fialkov13, Fialkov14b, Kaur2022, Magg22}. These models simulate vast cosmological volumes (e.g., boxes $\sim1$ Gpc across) with individual resolution elements that are a few Mpc across. Within these volume elements, the local SFRD is typically computed with integrals of the halo mass function \citep[e.g.][]{Jaacks18}, though recent work has been calibrated with merger tree-based semi-analytic models \citep{Magg22}. Utilizing our model to simulate variations within individual sub-volumes (as we present in Section \ref{sec:resultsIII}) can lay the groundwork to improve future sub-grid prescriptions within large-scale semi-numerical models.

\section{Fiducial Semi-Analytic Model} \label{sec:methods}
In this section, we introduce our fiducial semi-analytic model whose purpose is to self-consistently determine the evolution of the global SFRD (along with other aspects of early star formation such as the scatter in SFRD within sub-volumes of the Universe) given some input DM halo model, star formation prescription, and parameter values. We note that the efficiency of this framework allows each realization to be completed in $\sim$75 minutes on a single CPU. Our fiducial modelling parameters are summarized in Table \ref{tableOparams} and described in more detail below.

\subsection{Dark Matter Halo Merger History} \label{subsec:DMmethods}
Our  model uses MC merger trees based on EPS formalism \citep{Press74, EPSformalism} to model DM halo growth. We generate merger trees following the prescription in \cite{Lacey93}, which assumes binary mergers of DM halos at each redshift step, separated by $\Delta z$. For a given final halo mass and redshift, the merger history is modelled by iteratively stepping back through time, determining the first progenitor halo mass via the EPS mass-weighted progenitor function, and setting the second progenitor halo mass such that the sum of the two progenitors is equal to the merged descendent halo mass. This is done for each halo by solving equation 5 of \cite{Visbal14} for $\sigma_{\rm M_{1}}$, the root-mean-square (rms) density fluctuation on a scale corresponding to mass $\rm M_{1}$,
\begin{equation}
    \sigma_{\rm M_{1}} = \sqrt{\sigma_{\rm M_{0}}^{2} + \frac{1}{2} \left( \frac{\delta_{\rm crit}}{\erf^{-1}(x)} \right) ^{2} \left( \frac{1}{D(z+\Delta z)} - \frac{1}{D(z)} \right) ^{2}} .
\end{equation}
Here, $\sigma_{\rm M_{0}}$ is the rms density fluctuation for the scale corresponding to the present halo mass, $\delta_{\rm crit} = 1.686$ is the critical overdensity in linear perturbation theory, $D(z)$ is the linear growth factor, and $x$ is a randomly drawn value between 0--1. From $\sigma_{\rm M_{1}}$ we determine the progenitor halo mass, $M_{1}$, and the merging halo progenitor mass, $M_{2} = M_{0} - M_{1}$. This process is repeated at each step back through cosmic time until a halo falls below the resolution halo mass, $M_{\rm res}$, after which no further progenitors are modelled. To account for any statistically rare merger histories, we generate multiple merger trees for each halo mass bin whose masses are set at the end of the simulation, $z = 15$. 

In our fiducial model we use 36 logarithmically spaced mass bins from $10^{5.6}-10^{9.1}$ \Msun, finding this mass range to be sufficiently converged for estimating the global SFRD between $z = 15-35$. These precise mass values have little physical motivation, we simply find that adding halo mass bins above $10^{9.1}$ \Msun \ only affects the results at the earliest redshifts ($z \gtrsim 42$), and that the critical mass for PopIII star formation is higher than $10^{5.6}$ \Msun \ at $z = 15$ in all reported runs, indicating lower mass bins would not contribute to the resulting SFRDs. We also varied the number of halo mass bins and found that our fiducial number of bins yields SFRD values that are converged to within 10\% on average between $z = 15-35$. Within each mass bin we generate 10 merger trees each initialized at $z$ = 15, finding this number of trees to be the most reasonable trade-off between the required computation time and the statistical noise in our resulting SFRDs. We adopt a resolution halo mass of $M_{\rm res} = 5 \times 10^{4}$ \Msun \ for our merger trees, a threshold well below typical values of \Mcrit \ so that no early star formation is missed. We run our models from $z$ = 60--15, with redshift steps of $\Delta z$ = 0.05. Overall, we find the results of our work to be sufficiently converged for predicting the globally-averaged SFRD over the redshift range $z = 15-35$.

\subsection{Star Formation} \label{subsec:SFmethods}
Here we discuss the methods used to determine the formation of both PopIII and PopII stellar mass within merger tree halos. Note that we do not track individual stars in our model, and so any PopII and PopIII stellar masses discussed are total mass values. We focus mainly on the high-redshift transition from PopIII to PopII and leave any analysis of the transition from PopII to PopI for future work.

\subsubsection{Critical Mass \& PopIII Star Formation} \label{subsubsec:popIII}
We use a simple ``instantaneous" prescription for PopIII star formation in our fiducial model. A given halo without previous star formation will undergo a single episode of pristine star formation, creating a constant stellar mass of ${M_{\rm III,new}} = 200$ \Msun \ immediately after it surpasses the critical mass for star formation, \Mcrit. This is roughly consistent with the average PopIII stellar mass found within halos by \cite{Skinner20} (see their Figure 8), and approximately corresponds to a constant PopIII star formation efficiency of $f_{III} = 0.0001$ for a halo of mass $10^{6.1}$ \Msun. 

The critical mass is a global threshold for PopIII star formation, determined at each time step by finding the smaller of the atomic and molecular hydrogen cooling halo masses, i.e.
\begin{equation} \label{EQ-Mcrit}
    M_{\rm crit} = min(M_{\rm a}, M_{\rm H_{2}}) .
\end{equation}
Since the gas within early DM halos is pristine, molecular hydrogen (\Htwo) is initially the most efficient gas coolant. As star formation proceeds, the resulting radiative feedback dissociates \Htwo \ molecules, effectively raising the halo virial temperature and therefore $M_{\rm H_{2}}$. For halos with virial temperatures $T_{\rm vir} \ {\gtrsim 10^{4}}$ K, atomic hydrogen cooling becomes the more efficient mode of halo cooling, and so we adopt our atomic cooling mass from \cite{Visbal20},
\begin{equation} \label{EQ-Ma}
    M_{\rm a} = 5.4\times 10^{7} \ {\rm M_{\odot}} \left( \frac{1+z}{11} \right) ^{-1.5} .
\end{equation}
This mass is based on the threshold found in hydrodynamic cosmological simulations, with precise values from \cite{Fernandez14}. The \Htwo \ cooling mass used in our fiducial model is adopted from the hydrodynamical simulations of \cite{Kulkarni21}, which are the first \citep[along with][]{Schauer21} to include a dependency on both the relative baryon-DM streaming velocity, ${v_{\rm bc}}$, and the LW radiation background intensity, \JLW \ \citep[also see][]{Nebrin23}. They also consider an improved treatment of \Htwo \ self-shielding, ultimately leading to an analytic fitting function that is well-calibrated to the hydrodyamical simulations:
\begin{equation} \label{EQ-MH2}
    M_{\rm H_{2}} = M_{\rm z20}(J_{\rm LW}, v_{\rm bc}) \left( \frac{1+z}{21} \right) ^{\alpha (J_{\rm LW}, v_{\rm bc})}.
\end{equation}
Here, the critical mass at $z$ = 20, ${M_{\rm z20}}$, and the power index, $\alpha$, are both functions of the LW background intensity and baryon-DM streaming velocity (see \cite{Kulkarni21} for the best-fit values of these parameters to the simulations). At each time step, our semi-analytic model determines which halos have masses above \Mcrit \ and forms ${M_{\rm III,new}}$ within each. 

In Section \ref{subsec:Mcrit-results}, we compare the results of our fiducial model, with \Mmol \ determined by equation \ref{EQ-MH2}, to a previous model of \Mmol \ based on the simulations of \cite{Machacek01} \citep[also see][]{Stacy11, Greif11, Fialkov12}. This molecular mass fit, or variations thereof, has commonly been used in simulations of early star formation throughout the literature. Here we implement the form used in \cite{Visbal14} which includes a dependence on the redshift:
\begin{equation} \label{EQ-Machacek}
    M_{\rm H_{2}} = 2.5 \times 10^{5} \left( \frac{1+z}{26} \right)^{-1.5} \left(1 + 6.96 (4 \pi J_{\rm LW})^{0.47}\right) .
\end{equation}
The key difference between this and our fiducial equation for \Mmol \ is that the fit presented in \cite{Kulkarni21} includes the effects of \Htwo \ self-shielding and a stronger relative baryon-DM streaming velocity dependence. Fitting functions based on previous simulations, including those of \cite{Machacek01}, either do not include \vbc, or assumed that the LW intensity and \vbc \ were independent modelling processes and their effects on \Mcrit \ were multiplicative \citep{Fialkov12}. As shown by \cite{Kulkarni21}, their effects on \Mcrit \ are not multiplicative, hence our choice of this \Mmol \ fit for our fiducial model. 

\subsubsection{Delay Time for Enriched Star Formation} \label{subsubsec:t_delay}
A key parameter of our semi-analytic model is the delay time between the formation of PopIII and metal-enriched PopII stars, \tdelay. This delay period is a result of the radiative and supernovae feedback from the first stars which heats and ejects gas from within the hosting minihalos. Once the halo gas reservoir recollects and cools once more, the enriched material begins forming PopII stars \citep{Chiaki17, Chiaki13, Jeon14}. We choose \tdelay \ = 10 Myr for our fiducial model, and explore the impact of different \tdelay \ values on the resulting SFRDs in Section \ref{subsec:varyParams}.

\subsubsection{PopII Star Formation} \label{subsubsec:popII}
We assume that once PopIII stars form in a halo and the delay time has elapsed, it begins forming metal-enriched PopII stars. For PopII star formation, we adopt the system of differential equations described in \cite{Furlanetto21} that simultaneously determines the gas and stellar masses of a given halo through a simplified ``bathtub" model \citep[e.g.][]{Bouche10, Dave12, Dekel14}:
\begin{equation} \label{EQ-M*ODE}
    {\dot{M}}_{\rm *,II} = \frac{f_{\rm II}}{t_{\rm ff}} M_{\rm gas},
\end{equation}
\begin{equation} \label{EQ-MgasODE}
    {\dot{M}}_{\rm gas} = {\dot{M}}_{\rm acc} - {\dot{M}}_{\rm *,II} (1+\eta_{\rm SN}) .
\end{equation}
Here, ${\dot{M}}_{\rm *,II}$ is the halo PopII star formation rate, $f_{\rm II}$ is the PopII star formation efficiency per halo freefall time, $t_{\rm ff}$, which we assume to be 10\% of the Hubble time (i.e. $t_{\rm ff} = 0.1t_{\rm H}$), ${\dot{M}}_{\rm gas}$ represents the time derivative of the halo gas mass, ${\dot{M}}_{\rm acc}$ is the mass accretion rate of the halo, and $\eta_{\rm SN}$ is the supernova ejection efficiency (described in more detail below). By analytically solving these differential equations at each time step, we obtain the gas and stellar masses of each halo, and determine the gas mass ejected through SNe by subtracting these two masses from the initial total baryonic mass of the halo. This prescription forms stars such that a fraction of the total halo gas mass, $f_{\rm II}$, is converted into stellar mass over one halo freefall time, allowing isolated DM halos to continue forming PopII stars as long as there is gas within them.

While we note that there is large uncertainty in the star formation efficiency (SFE) of unobserved low-mass galaxies at very high redshifts ($z \gtrsim 15$), we have aimed to select a realistic value for our fiducial choice of $f_{\rm II}=0.0025$. When combined with the SNe feedback prescription described below, we find that this choice of $f_{\rm II}$ leads to a SFRD consistent with \cite{Visbal18}, which was calibrated to observations of $z{\sim}6$ UV galaxy luminosity functions \citep{2015ApJ...803...34B}. In a related upcoming study (Hazlett et al., in prep.), we have created a model of PopII star formation directly calibrated to the Renaissance simulations. We find that this calibration yields PopII SFRDs that agree to within a factor of two at $z \lesssim 32$ with those predicted in our fiducial model presented here.

At each time step, we determine the PopII and PopIII star formation rates (SFRs) for each merger tree by dividing any new stellar mass formed by the length of the time step. We average the SFRs of all merger trees in each mass bin, and weight each average by the corresponding halo mass bin number density, $n_{i}$, as determined from the Sheth-Tormen halo mass function \citep{ShethTormen} at bin center, i.e.
\begin{equation} \label{EQ-N_halos}
    n_{i} = \int_{m_{\rm l}}^{m_{\rm u}} \frac{dn}{dM} dM .
\end{equation}
Here, $dn/dM$ is the Sheth-Tormen halo mass function and $n_{i}$ is the number density of halos at $z = 15$ in mass bin $i$, bounded by upper and lower halo masses $m_{\rm u}$ and $m_{\rm l}$. For the remainder of this paper, all DM halo mass bin number densities are the $z = 15$ values unless otherwise stated. The weighted averages of all mass bins are then summed to give the global PopII and PopIII SFRD values for that redshift step, i.e.
\begin{equation} \label{EQ-SFRD}
    SFRD(z) = \sum_{i}n_{i}{{\langle {\dot{M}}_{*,\rm tot} \rangle}_{i}} .
\end{equation}
Here, ${\dot{M}}_{\rm *,tot}$ is the total SFR of every halo at redshift $z$ within a merger tree in mass bin $i$. Averaging the total SFRs of all merger trees in mass bin $i$ gives ${{\langle {\dot{M}}_{\rm *,tot} \rangle}_{i}}$ which we then weight by $n_{i}$ and sum across all mass bins.

\subsection{Feedback Processes} \label{subsec:Feedbackmethods}
Our semi-analytical framework also includes various crucial feedback processes that impact global SFRD evolution. In this subsection, we will discuss the auxiliary features and physical feedback processes of our fiducial model and how they affect early star formation.

\subsubsection{Lyman-Werner Background Intensity} \label{subsubsec:LW}
LW feedback must be considered in any semi-analytic model of early star formation. As the first stars begin to shine and emit radiation, photons in the energy range ${E_{\rm LW}}$ = 11.2--13.6 eV freely stream out up to a horizon of $\sim$100 Mpc and into other DM halos, dissociating \Htwo \ molecules within them \citep{Haiman97, Haiman00, Machacek01, Oshea08, Ahn09}. In our fiducial model, we self-consistently determine the LW background intensity at each time step in units of $10^{-21} \rm erg \ s^{-1} cm^{-2} Hz^{-1} sr^{-1}$ using the \JLW \ calculation from \cite{Visbal20}:
\begin{equation} \label{EQ-LW}
    J_{\rm LW}(z) = \frac{c(1+z)^{3}}{4\pi} \int_{z}^{\infty} \epsilon_{\rm LW}({z^{\prime}}) \Bigg|\frac{dt_{\rm H}}{dz^{\prime}}\Bigg| f_{\rm LW}(z^{\prime}, z) dz^{\prime} .
\end{equation}
Here, $\epsilon_{\rm LW}({z^{\prime}})$ is the mean LW emissivity, $t_{\rm H}$ is the Hubble time, and $f_{\rm LW}(z^{\prime}, z)$ is the attenuation of the LW flux from $\rm {z^{\prime}}$ to $z$ as LW photons redshift into Lyman series lines and are absorbed \citep{Haiman97}. We approximate this LW attenuation using equation 22 in \cite{Ahn09}, and determine $\epsilon_{\rm LW}({z^{\prime}})$ by summing the contributions of both stellar populations. 

We assume the number of LW photons produced per stellar baryon to be $\eta_{\rm III}$ = 65000 for PopIII stars \citep{Schaerer02}, and $\eta_{\rm II}$ = 4000 for metal-enriched stars \citep{Samui07, Incatasciato23}. We set these to be equal to the number of ionizing photons produced per stellar baryon for simplicity, and also note that our assumed value of $\eta_{\rm III}$ may be an overestimate, but we have found that the PopII SFRD quickly dominates the LW background in all runs, and varying $\eta_{\rm III}$ only changes the PopIII SFRD by a few percent. An increase in \JLW \ results in an increase in both ${M_{\rm z20}}$ and $\alpha$ in equation \ref{EQ-MH2}, giving a larger \Mmol \ for equation \ref{EQ-Mcrit}. Increased LW radiation intensity therefore results in an overall increase to the critical mass for PopIII star formation, suppressing further star formation.

In our model, we do not consider the effects of X-ray feedback on PopIII star formation. We note that \cite{Hegde23} recently found that the X-ray background is not a dominant effect for PopIII star formation in minihalos at the redshifts considered in this work \citep[$z>15$, also see][]{Ricotti16, Ricotti04}.

\begin{table*}[t]
    \centering
    \begin{tabular}{c c c c}
        \hline
        Parameter & Description & Fiducial Value & Reference\\
        \hline\hline
        \sigvbc & Relative baryon-DM streaming velocity & 30 km $\rm s^{-1}$ & \cite{v_bcRef} \\
        $\eta_{\rm III}$ & \# of LW photons per PopIII baryon & 65000 & \cite{Schaerer02}\\
        $\eta_{\rm II}$ & \# of LW photons per PopII baryon & 4000 & \cite{Samui07} \\
        $M_{\rm H_{2}}$ & Molecular hydrogen cooling mass & $M_{\rm z20}(J_{\rm LW}, v_{\rm bc}) \left( \frac{1+z}{21} \right) ^{\alpha (J_{\rm LW}, v_{\rm bc})}$ & \cite{Kulkarni21}\\
        $M_{\rm a}$ & Atomic hydrogen cooling mass & $5.4\times 10^{7} {\rm M_{\odot}} \left( \frac{1+z}{11} \right) ^{-1.5}$ & \cite{Visbal20} \\
        $M_{\rm III,new}$ & PopIII formation mass & 200 \Msun & \cite{Skinner20} \\
        $t_{\rm delay}$ & Delay time before PopII star formation & 10 Myr & \cite{Jeon14} \\
        $f_{\rm II}$ & Enriched star formation efficiency & 0.0025 & see Section \ref{subsubsec:popII} in main text \\
        $\eta_{\rm SN}$ & Supernova ejection efficiency & $5.23 \times \left( \frac{v_{\rm esc}}{14.46 \ \rm km \ \rm s^{-1}} \right)^{-2}$ & \cite{Sassano21}\\
        \\
        \hline
    \end{tabular}
    \caption{Parameters and fiducial values adopted in our framework for modelling the global SFRD.}
    \label{tableOparams}
\end{table*}

\subsubsection{Relative Baryon-DM Streaming Velocity} \label{subsubsec:vbc}
Another important parameter in modelling early star formation is the relative streaming velocity between baryonic and dark matter \citep{v_bcRef, Tseliakhovich11, Fialkov13, Greif11, Stacy12}. At cosmic recombination, photons decoupled from the baryonic matter which was then free to interact gravitationally. Dark matter density perturbations, however, could grow under the influence of gravity before this. As a result, we are left with a Maxwell-Boltzmann distribution of relative baryon-DM streaming velocities with a rms value of \sigvbc \ = 30 km $\rm s^{-1}$ at recombination, which decreases with redshift as $v_{\rm bc}(z) = v_{\rm bc,0} ((1+z)/1100)$. This relative motion is roughly constant over scales of ${\sim}3$ Mpc, and is crucial for modelling early star formation because the DM halos are unable to efficiently capture and cool the high-speed gas streaming by them. An increase in streaming velocity leads to a decrease in the amount of gas bound to halos and delays halo gas cooling, thus leading to an overall increase in \Mcrit. The value of this velocity is commonly referred to in terms of \sigvbc, and while we explore the impact of various \sigvbc \ values below, we adopt a fiducial global streaming velocity of \vbc \ = 1\sigvbc \ = 30 km $\rm s^{-1}$. 

\subsubsection{Feedback from Supernovae} \label{subsubsec:SNe}
A fraction of the gas mass within a given halo will be ejected by the star formation and SNe occurring at a given time step. The gas mass ejected by PopII SNe is determined by the SN ejection efficiency of the halo, $\eta_{\rm SN}$. In this work, we follow a similar prescription for determining $\eta_{\rm SN}$ to the one presented in \cite{Sassano21}:
\begin{equation} \label{EQ:eta_SN}
    \eta_{\rm SN} = \frac{2 E_{\rm SN} \epsilon_{\rm SN} R_{\rm SN}}{v_{\rm esc}^{2}} ,
\end{equation}
where $E_{\rm SN}$ is the average energy per PopII supernova, $\epsilon_{\rm SN}$ is an efficiency parameter, $R_{\rm SN}$ is the rate of SNe per solar mass, and $v_{\rm esc}$ is the escape velocity of the halo, which we assume to be the halo circular velocity determined by equation 25 of \cite{Barkana01}. We note that the circular and escape velocities at a halo's virial radius differ by a factor of $\sqrt{2}$, however the effect that SNe ejection has on the global SFRD is comparatively low (i.e. removing all SNe feedback results in a $\sim30 \%$ increase to the $z=15$ PopII SFRD), and since we do not assume a preferred radius for SNe ejection from our DM halos, we adopt the circular velocity for simplicity. We adopt values of $E_{\rm SN} = 10^{51}$ ergs and $\epsilon_{\rm SN} =$ 0.0016 from \cite{Wise12}. $R_{\rm SN}$ is determined by dividing the SN energy per solar mass, $E_{\rm SN, M_{\rm \odot}}$, by $E_{\rm SN}$. Adopting a value of $E_{\rm SN, M_{\rm \odot}} = 6.8 \times 10^{48}$ ergs $\rm M_{\odot}^{-1}$ \citep{Wise12} gives us $R_{\rm SN} = 0.0068 \ \rm M_{\odot}^{-1}$. The numerator of equation \ref{EQ:eta_SN} can then be described in terms of velocity, and so we normalize this expression for a halo at $z = 20$ with a mass equal to the atomic cooling mass (equation \ref{EQ-Ma}). This gives us $\eta_{\rm SN} = 5.23 \times (v_{\rm esc} / 14.46 \ \rm km \ s^{-1})^{-2}$.

The ejected gas mass, $M_{\rm ej}$, is determined at each time step by subtracting the final stellar and gas masses within the halo from its initial total baryonic mass. $M_{\rm ej}$ is then removed from the available halo gas mass until it is reintroduced for future star formation after one freefall time, which we have assumed to be 10\% of the Hubble time at the moment of ejection.

\section{Results I: Fiducial Model} \label{sec:resultsI}
In this section we present the global SFRDs of our fiducial model. We illustrate the effects of a new critical mass model (equation \ref{EQ-MH2}) that more accurately considers the effects of LW feedback, \vbc, and \Htwo \ self-shielding, as well as the impact of a halo-to-halo scatter on \Mcrit. We also assess the impact that various physical processes and parameter value choices have on the results. Unless otherwise stated, all SFRD values plotted throughout this work have been smoothed post-simulation with a running average over $\Delta z = 1$. We also provide analytic fits for a sample of the LW backgrounds that result from this research in Appendix \ref{appendixLW}.

\begin{figure*}
    \centering
    \includegraphics[width=\textwidth]{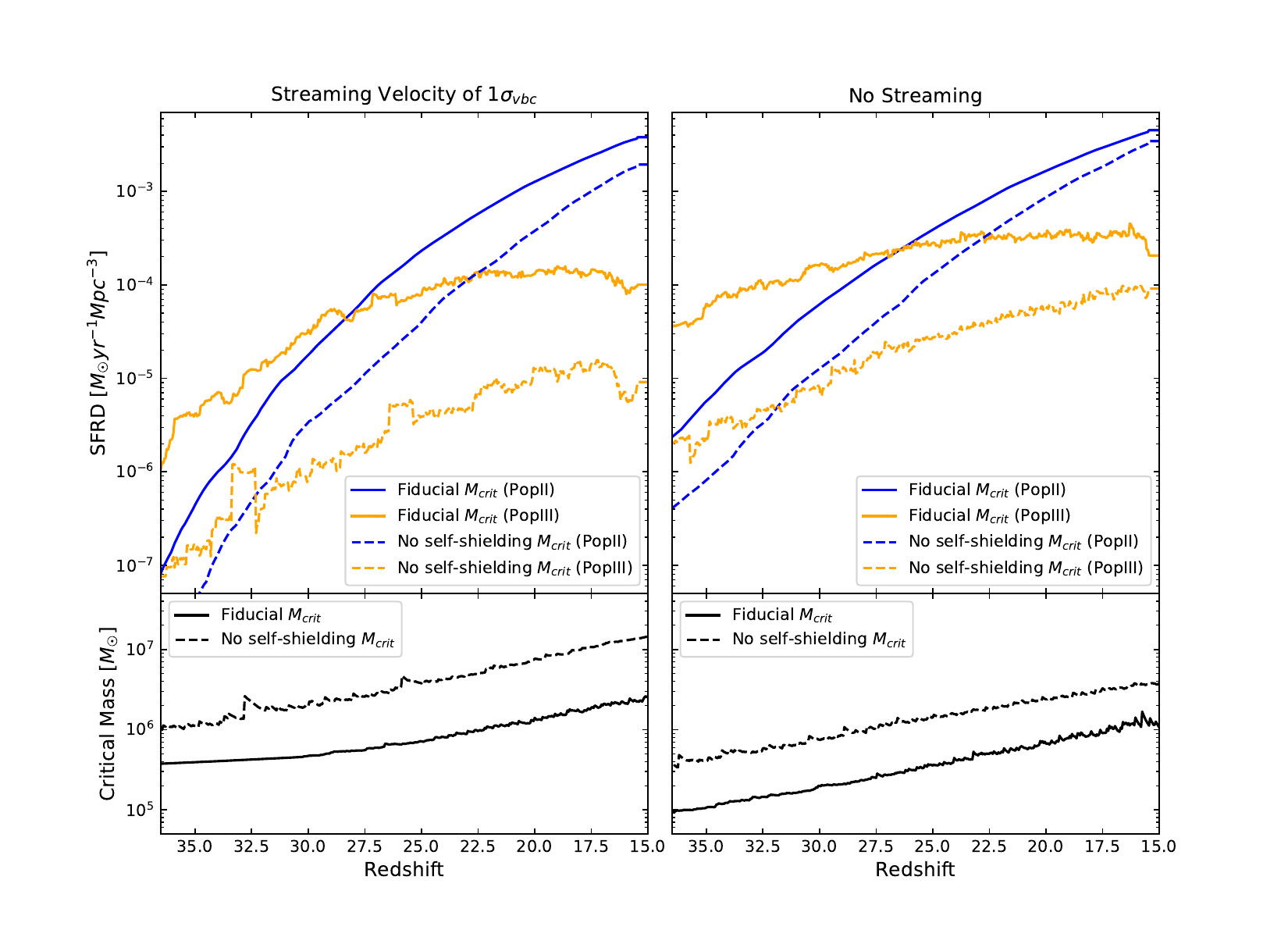}
    \caption{The impact of different critical mass calculation methods for realizations including (left column) and not including (right column) relative baryon-DM streaming. \textbf{Top left:} Comparison of the PopII (blue) and PopIII (orange) SFRDs with different methods for determining \Mcrit \ including a global streaming velocity of 1$\sigma$ = 30 km $s^{-1}$. Our fiducial framework using the \cite{Kulkarni21} \Mcrit \ model (solid) is compared to a realization using the critical mass model from \cite{Visbal20}, based on the simulations of \cite{Machacek01} (dashed). \textbf{Top right:} Same as top left, but for realizations with no global baryon-DM streaming (i.e. $v_{\rm bc}$ = 0 km $s^{-1}$). \textbf{Bottom left:} The value of \Mcrit ($z$) for each method with a streaming velocity of 1\sigvbc. \textbf{Bottom right:} Same as bottom left but with no baryon-DM streaming.}
    \label{fig:Mcrit-model}
\end{figure*}

\subsection{Critical Mass Model} \label{subsec:Mcrit-results}
Figure \ref{fig:Mcrit-model} shows the SFRDs and corresponding critical masses resulting from both our fiducial model with \Mcrit \ based on the new simulations of \cite{Kulkarni21}, and those resulting from the \cite{Visbal20} adaptation of the commonly used critical mass model consistent with the simulations of \cite{Machacek01}, \cite{Greif11}, and \cite{Stacy12} in which the effects of \Htwo \ self-shielding are not considered. This ``no self-shielding" \Mcrit \ model stems from the work of \cite{Fialkov12}, which updated the critical mass threshold from \cite{Machacek01} to include more sophisticated treatments for \vbc -dependence (also see equation 3 of \cite{Visbal20} and accompanying text for more information on including LW and \vbc -dependence). However, since recent works have revised the critical mass threshold for PopIII star formation to include the effects of \Htwo \ self-shielding \citep[e.g.][]{Kulkarni21, Schauer21, Nebrin23}, we refer to this earlier \Mcrit \ as the ``no self-shielding" critical mass model for the remainder of this paper.

The left set of plots is for a global streaming velocity of 1\sigvbc. Here we see that our fiducial critical mass is a factor of a few lower than the no self-shielding model throughout, falling to a factor of $\sim$6 times lower at $z\sim$ 15. This lower \Mcrit \ leads to earlier star formation and an overall larger number of star-forming halos in the fiducial model. We thus find a PopIII SFRD that is over an order of magnitude higher than the no self-shielding PopIII SFRD throughout most of the redshifts shown. The fiducial PopII SFRD is also consistently higher than the one given by the no self-shielding \Mcrit, but this difference falls from an order of magnitude at $z=35$ to a factor of two times the no self-shielding PopII SFRD at $z=15$.

The right set of plots in Figure \ref{fig:Mcrit-model} show the results for no baryon-DM streaming. Here we see a somewhat smaller difference in the SFRDs. At $z = 35$, our fiducial model gives SFRD values that are factors of 30 and 8 times higher than the no self-shielding values for PopIII and PopII, respectively. The final PopII SFRD of the fiducial model is only 30\% higher than the no self-shielding case, whereas the final PopIII SFRD is a factor of $\sim$2 higher. Here, the effects that cause the differences in the 1\sigvbc \ case are reduced because the critical masses fall within a factor of five to one another throughout, owing to the stronger \vbc \ dependence in the no self-shielding model.

\begin{figure}
    \centering
    \includegraphics[width=0.5\textwidth]{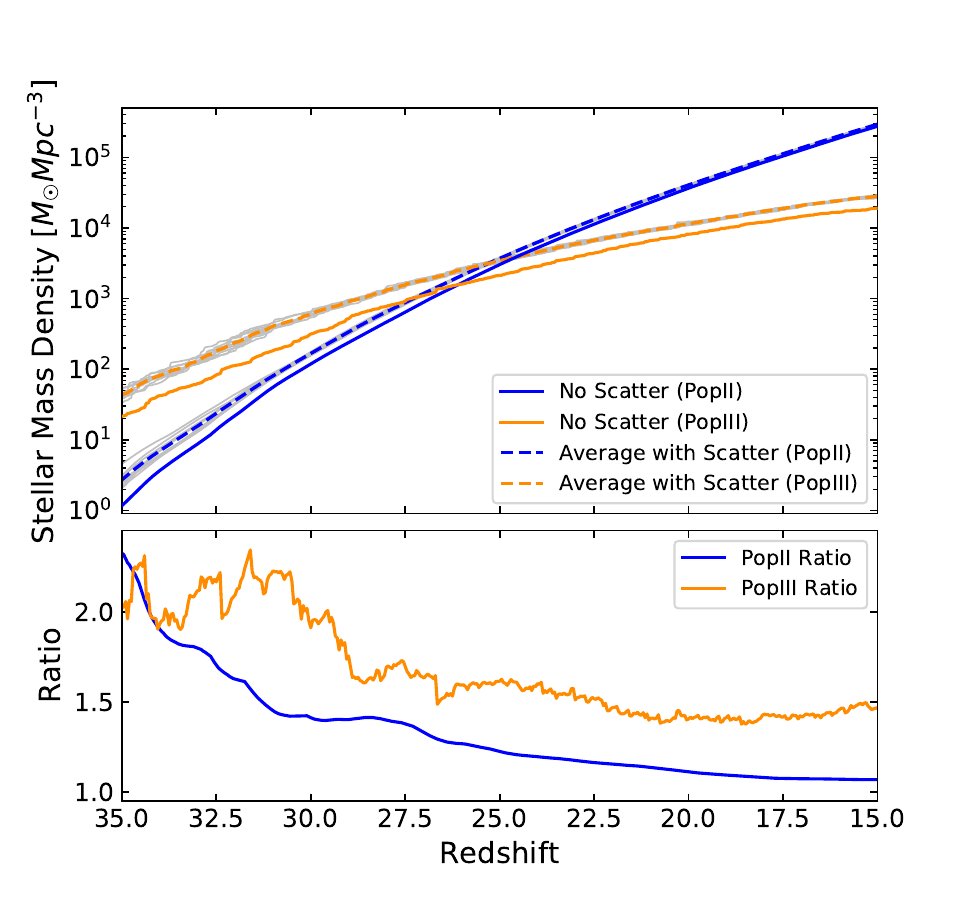}
    \caption{The impact of a critical mass scatter on the results of our fiducial model. \textbf{Top:} The evolution of the stellar mass density over time. The fiducial PopII (blue) and PopIII (orange) SMDs without scatter are shown by the solid lines. The densities for the sample of ten runs including \Mcrit \ scatter are plotted in gray, and the average of these ten runs for both stellar populations are shown by the corresponding dashed lines. \textbf{Bottom:} Ratio of the average SMD including \Mcrit \ scatter to the fiducial model SMD for each stellar population.}
    \label{fig:Scatter}
\end{figure}

\subsection{Critical Mass Scatter} \label{subsec:scatter}
In our fiducial model we assume \Mcrit \ to be a global constant at each time step; we now consider the impact of a scatter on \Mcrit($z$) which effectively gives each DM halo an individual critical mass threshold for PopIII star formation, $M_{\rm crit,halo}$. To more accurately account for variations in halo mass assembly and geometry, \cite{Kulkarni21} quantified this scatter in terms of \JLW \ and \vbc \ which we may implement via
\begin{equation} \label{EQ-scatter}
    \log_{10}(M_{\rm crit,halo}) = \log_{10}(M_{\rm crit}) + R .
\end{equation}
Here, the value of $R$ is determined for each DM halo after every halo freefall time by randomly drawing from a Gaussian distribution centered on zero and with a standard deviation of $\sigma =$ 0.15. The \Mcrit \ scatter values reported in Figure 7 of \cite{Kulkarni21} represent the difference between the $25^{\rm th}$ and $75^{\rm th}$ percentile of critical masses, equivalent to $1.34\sigma$. For a streaming velocity of 1\sigvbc, this value ranges from 0.15--0.3, and we adopt the value 0.2, roughly corresponding a LW background intensity of \JLW = $1 \times 10^{-21} \rm erg \ s^{-1} cm^{-2} Hz^{-1} sr^{-1}$. Although we self-consistently determine \JLW \ at each time step, its values at $z \lesssim$ 30 are comparable to unity, and so we conservatively adopt $\sigma = $ 0.15 ($ \simeq 0.2 / 1.34$) for the standard deviation of our distribution. \citep[for more information, see Figure 7 of][]{Kulkarni21}. 

We illustrate the impact of this relative scatter on \Mcrit \ in Figure \ref{fig:Scatter}. The solid lines in the top panel show the stellar mass densities (SMDs) for both PopII and PopIII as determined by our fiducial model. The SMDs of ten simulations including \Mcrit \ scatter are shown in light gray for each stellar population, and the average of these ten SMDs are represented by the corresponding dashed lines. We choose to show the SMDs in Figure \ref{fig:Scatter} rather than the SFRDs to more clearly show the impact that \Mcrit \ scatter has on the star formation resulting from our semi-analytic model.

We see from the bottom panel of Figure \ref{fig:Scatter} that introducing \Mcrit \ scatter increases the PopIII SMD by a factor of $\sim$2 at early times ($z \gtrsim$ 30). Below $z\sim30$, the average PopIII SMD with scatter maintains a roughly 50\% increase over the no scatter case. This increase is due to halos drawing low random $R$ values for equation \ref{EQ-scatter}, resulting in halo critical masses lower than the global mean value and earlier star formation than without scatter. The average PopII SMD with scatter is also increased by a factor of $\sim$2 at the highest redshifts shown, before tapering off and coming into agreement with the no scatter SMD at the percent level by $z$ = 15. PopIII star formation relies solely on a halo overcoming \Mcrit, and halos are much more likely to randomly draw a low enough $R$ value to cause earlier star formation than they are to consistently draw high $R$ values that keep $M_{\rm crit,halo} >$ \Mcrit. We therefore see a consistent increase in the PopIII SMD that is not strongly reflected by the PopII as the PopII SFRD does not depend as directly on \Mcrit.

\begin{figure}
    \centering
    \includegraphics[width=0.5\textwidth]{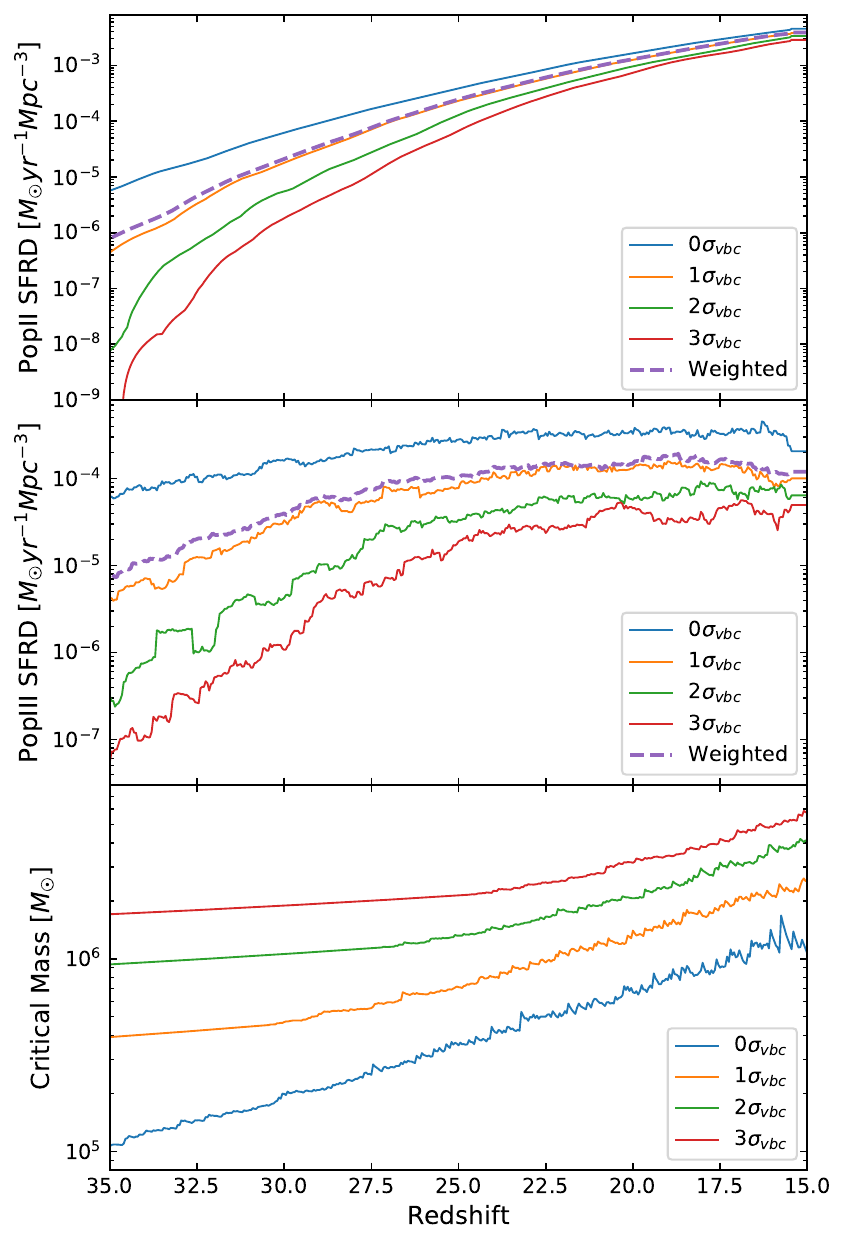}
    \caption{The impact of the relative baryon-dark matter streaming velocity on the global SFRD and critical mass. \textbf{Top:} PopII SFRDs for four models with global streaming velocities of 0$\sigma_{\rm vbc}$ = 0 $\rm km \ s^{-1}$ (blue), 1$\sigma_{\rm vbc}$ = 30 $\rm km \ s^{-1}$ (orange), 2$\sigma_{\rm vbc}$ = 60 $\rm km \ s^{-1}$ (green), 3$\sigma_{\rm vbc}$ = 90 $\rm km \ s^{-1}$ (red). The average PopII SFRD is also shown (purple dashed), with each SFRD weighted by the probabilities of the streaming velocity used, which is set by a Maxwell-Boltzmann distribution. \textbf{Middle:} Same as the top panel, but for the PopIII SFRDs. \textbf{Bottom:} The value of \Mcrit \ over time for each of the constant velocity runs. The close agreement between the weighted average SFRDs and those of the 1$\sigma_{\rm vbc}$ case justifies our fiducial streaming velocity of 30 $\rm km \ s^{-1}$.}
    \label{fig:vbc_Streaming}
\end{figure}

\begin{figure*}
    \centering
    \includegraphics[width=\textwidth]{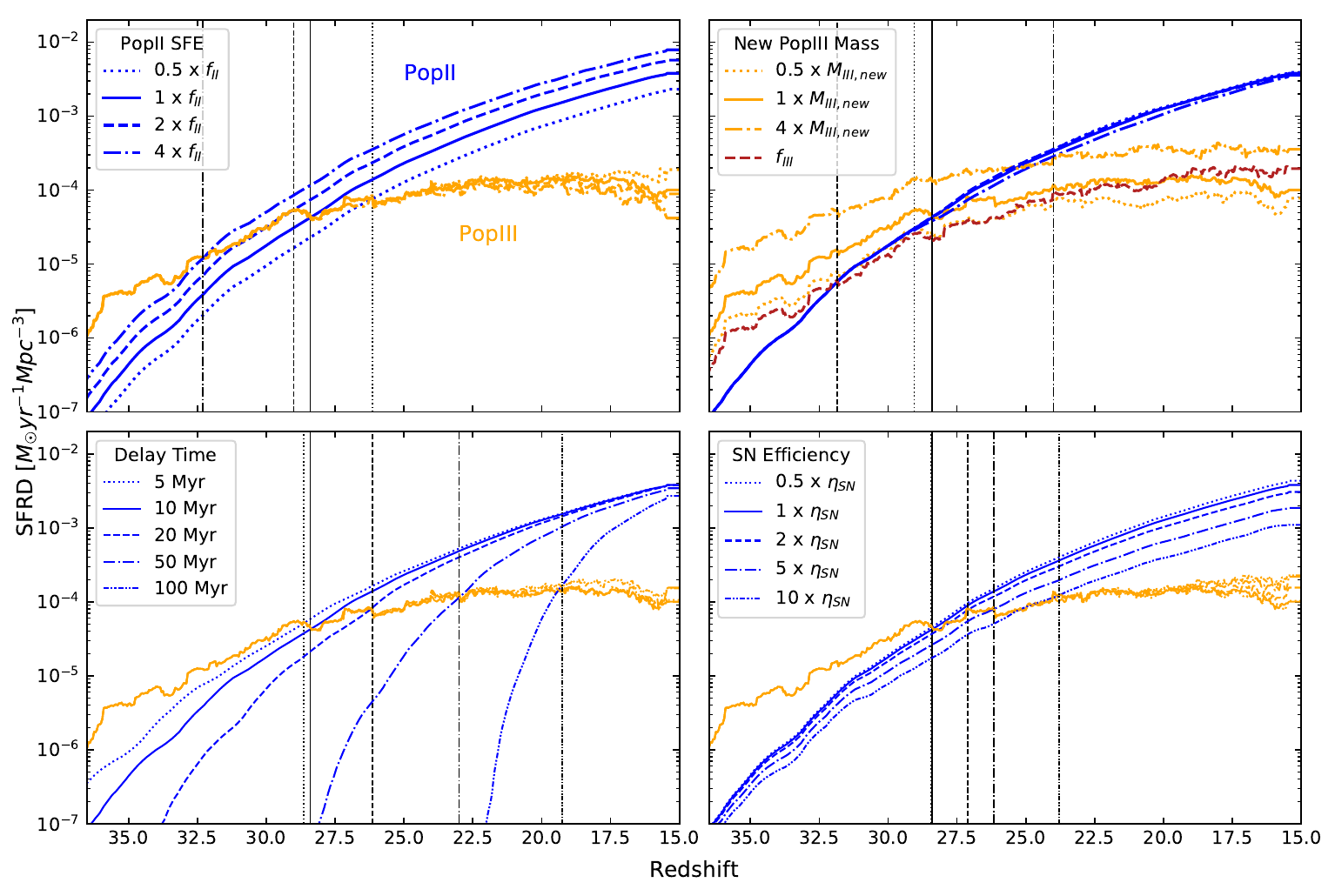}
    \caption{The effect of varying key model parameters on the global PopII (blue) and PopIII (orange) SFRDs. \textbf{Top left:} The impact of varying PopII SFEs on the global SFRD evolution. We vary our fiducial SFE (solid) by multiplying $f_{\rm II}$ by factors of 0.5 (dotted), 2 (dashed), and 4 (dot-dashed). The redshift at which the PopII SFRD first dominates over the PopIII SFRD in each of these simulations is identified by the corresponding vertical black lines in all panels. \textbf{Top right:} Similarly, the SFRD impact of varying the PopIII formation mass, $M_{\rm III,new}$. We multiply our fiducial $M_{\rm III,new}$ value by factors of 0.5 (dotted) and 4 (dot-dashed), and also include the SFRDs resulting from a constant PopIII SFE of $f_{\rm III} = 0.001$ (dashed dark orange, for clarity). \textbf{Bottom left:} SFRDs for five different values of \tdelay: 5 Myr (dotted), 10 Myr (solid, fiducial), 20 Myr (dashed), 50 Myr (dot-dashed), and 100 Myr (dot-dot-dashed). \textbf{Bottom right:} SFRDs for different supernova gas ejection efficiencies: 0.5$\eta_{\rm SN}$ (dotted), $\eta_{\rm SN}$ (solid, fiducial), 2$\eta_{\rm SN}$ (dashed), 5$\eta_{\rm SN}$ (dot-dashed), and 10$\eta_{\rm SN}$ (dot-dot-dashed).}
    \label{fig:Parameters}
\end{figure*}

\subsection{Baryon-Dark Matter Streaming} \label{subsec:streaming}
In Figure \ref{fig:vbc_Streaming}, we present PopII and PopIII SFRDs and their corresponding \Mcrit \ values for global streaming velocities of 0--3\sigvbc \ (0--90 km $\rm s^{-1}$). Recall that the velocities stated are those at recombination, which fall off with decreasing redshift. We also show the average SFRD determined by using many global streaming velocities ranging from 0--3\sigvbc \ and weighting each by the probability of the \vbc \ value used. The probabilities are determined from the Maxwell-Boltzmann distribution of \vbc \ values found at recombination \citep{Fialkov12, Tseliakhovich11, v_bcRef}, and these weighted averages are shown by the dashed purple curves in the top and middle panels of Figure \ref{fig:vbc_Streaming} for PopII and PopIII SFRDs, respectively.

We see from the top two panels of Figure \ref{fig:vbc_Streaming} that the SFRDs of both stellar populations are increasingly suppressed with faster streaming velocities. This is because higher \vbc \ values increase \Mcrit, which delays any PopIII and subsequent PopII star formation. This effect is greater for PopIII since it is directly governed by \Mcrit, whereas PopII star formation begins after \tdelay \ and depends on the gas mass of the host halo thereafter. The weighted average SFRDs each agree with those of the fiducial 1\sigvbc \ case to within 50$\%$ at $z < $ 35, indicating that our fiducial treatment is a reasonable representation of the global streaming velocity. We note that the true baryon-DM streaming velocity varies over spatial distances of a few Mpc, a detail that we plan to incorporate into future work.

\subsection{Additional Parameter Variations} \label{subsec:varyParams}
In this subsection, we explore the effect of various parameter value choices on the resulting SFRDs. To show the impact of our fiducial choices for the PopII SFE and PopIII formation mass, we vary the values of $f_{\rm II}$ and $M_{\rm III,new}$ and plot the resulting SFRDs in the top row of Figure \ref{fig:Parameters}. 
The top left panel shows the SFRDs for four realizations of our semi-analytic framework with PopII SFEs ranging from $f_{\rm II}$ = 0.00125--0.01. From here we see that the SFRD response is less than linearly proportional to changes in $f_{\rm II}$. Each PopII SFRD shown falls within a factor of four to one another at $z \lesssim$ 22, despite the value of $f_{\rm II}$ varying by a factor of eight. Any changes in the PopIII SFRDs ($z \lesssim$ 30) are only caused by differences in the PopII SFRDs altering the LW background intensity and thus \Mcrit. In each panel, we show the redshift at which the PopII SFRD first surpasses the PopIII by the corresponding vertical black lines. We see that higher $f_{\rm II}$ values cause the PopII SFRD to first dominate over the PopIII at earlier times, ranging from $z \approx 27$ for $0.5f_{\rm II}$ to $z \approx 32$ for $4f_{\rm II}$.

Looking to the top right panel of Figure \ref{fig:Parameters}, we see the SFRDs for three realizations with varying PopIII formation mass values ranging from $M_{\rm III,new} =$ 100--800 \Msun, as well as a realization with a constant PopIII SFE of $f_{\rm III} = 0.001$ which instantly converts a fraction $f_{\rm III}$ of a given halo's gas mass into PopIII stellar mass. The PopIII SFRDs with constant stellar mass each fall within a factor of five to one another at $z \lesssim$ 30 despite $M_{\rm III,new}$ changing by a factor of eight, meaning that we again find a less than linearly proportional response to varying $M_{\rm III,new}$. Using $f_{\rm III}$ instead gives slightly steeper growth in the PopIII SFRD, but it still falls within a factor of four to the fiducial model throughout the redshifts shown. Since the stellar mass formed in this realization is dependent on halo gas mass, the SFRD more strongly reflects halo mass growth over time. As with $f_{\rm II}$, higher PopIII SFRDs experience increased LW feedback which boosts \Mcrit \ and suppresses further star formation. The PopII SFRDs, however, are virtually unchanged by the varying PopIII star formation prescriptions since enriched star formation is mainly governed by the onset of PopIII star formation and \tdelay. 

The \tdelay \ parameter represents the delay between PopIII and metal-enriched star formation resulting from feedback associated with star formation and SNe winds. To show the impact of this, we vary our fiducial delay time and present the resulting global SFRDs in the bottom left panel of Figure \ref{fig:Parameters}. Here we see the SFRDs of five different time delays ranging from 5--100 Myr. Unsurprisingly, changes in \tdelay \ mostly affect the timing of the initial PopII star formation event. Regardless of the length of the delay, however, enriched star formation quickly overtakes the PopIII SFRD afterwards and gives similar final PopII SFRD values, all agreeing to within 75\% at $z = 15$. 

To show the effect of the supernova ejection efficiency, we vary the fiducial value of $\eta_{\rm SN}$ in our model and present the global SFRDs in the bottom right panel of Figure \ref{fig:Parameters}. Here we see that $\eta_{\rm SN}$ mostly affects the PopII SFRD at late times, with lower efficiencies yielding higher PopII SFRDs overall. At redshift 15, the 0.5$\eta_{\rm SN}$ case is a factor of $\sim$5 higher than the 10$\eta_{\rm SN}$ case. The effect of supernovae feedback increases with time, partially due to the redshift dependence of $v_{\rm esc}$ in equation \ref{EQ:eta_SN} causing more gas to escape over cosmic time.

\section{Results II: Impact of Enriched Star Formation \& DM Halo Models} \label{sec:resultsII}
In this section, we address the SFRD changes that result from alternative methods for determining the DM halo mass evolution and enriched star formation. We present a direct comparison of the SFRDs resulting from MC merger trees and smooth accretion models of DM halo evolution \citep[as in][]{Furlanetto17}. We also compare the results of our fiducial enriched star formation prescription (equations \ref{EQ-M*ODE} \& \ref{EQ-MgasODE}) to those of a simpler instantaneous model of star formation \citep[e.g.][]{Sun16, Park19, Magg18, Magg22}. For completeness, we also include an analytic approach reliant on halo mass function integration to estimate the SFRD \citep[e.g.][]{Munoz23, Munoz22, Mashian16, Visbal15a}. 

\subsection{Descriptions of Alternative Models} \label{subsec:AltModels}

\begin{figure}
    \centering
    \includegraphics[width=0.45\textwidth]{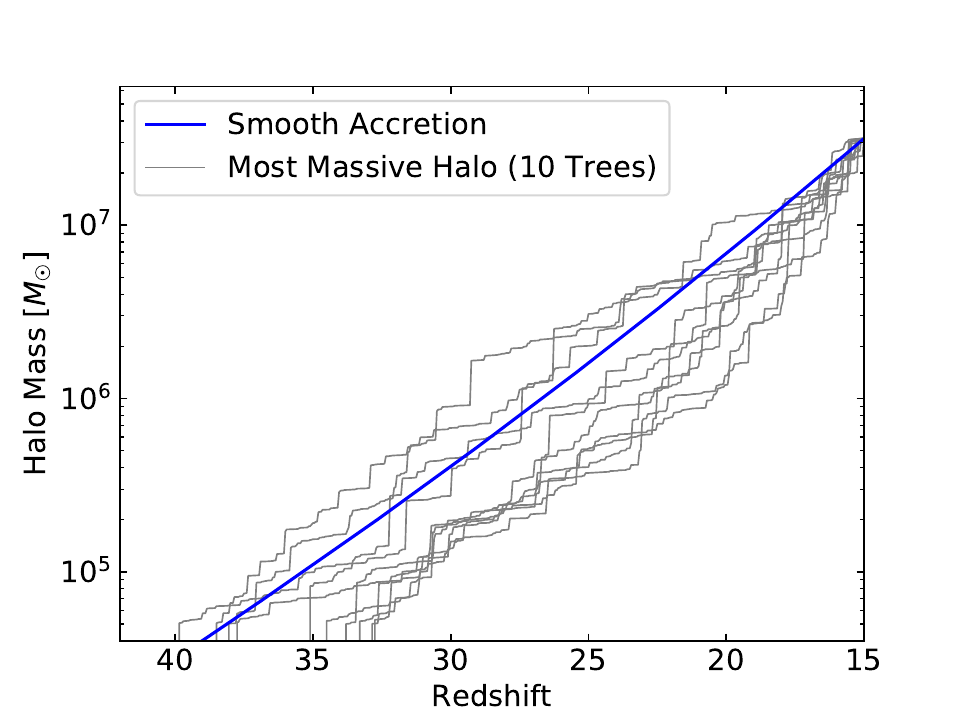}
    \caption{Comparison of mass growth histories for DM halos with a final masses of $\log_{10}(M_{\rm halo}/{M_{\odot}})$ = 7.5 at $z=15$. We show both the MC EPS approach implemented in our fiducial model (gray curves) as well as the smooth accretion method (blue curve) from \cite{Furlanetto17} (each described in detail in Sections \ref{sec:methods} and \ref{sec:resultsII}, respectively).
    }
    \label{fig:SA_vs_MC_Halos}
\end{figure}

\begin{figure*}
    \centering
    \includegraphics[width=\textwidth]{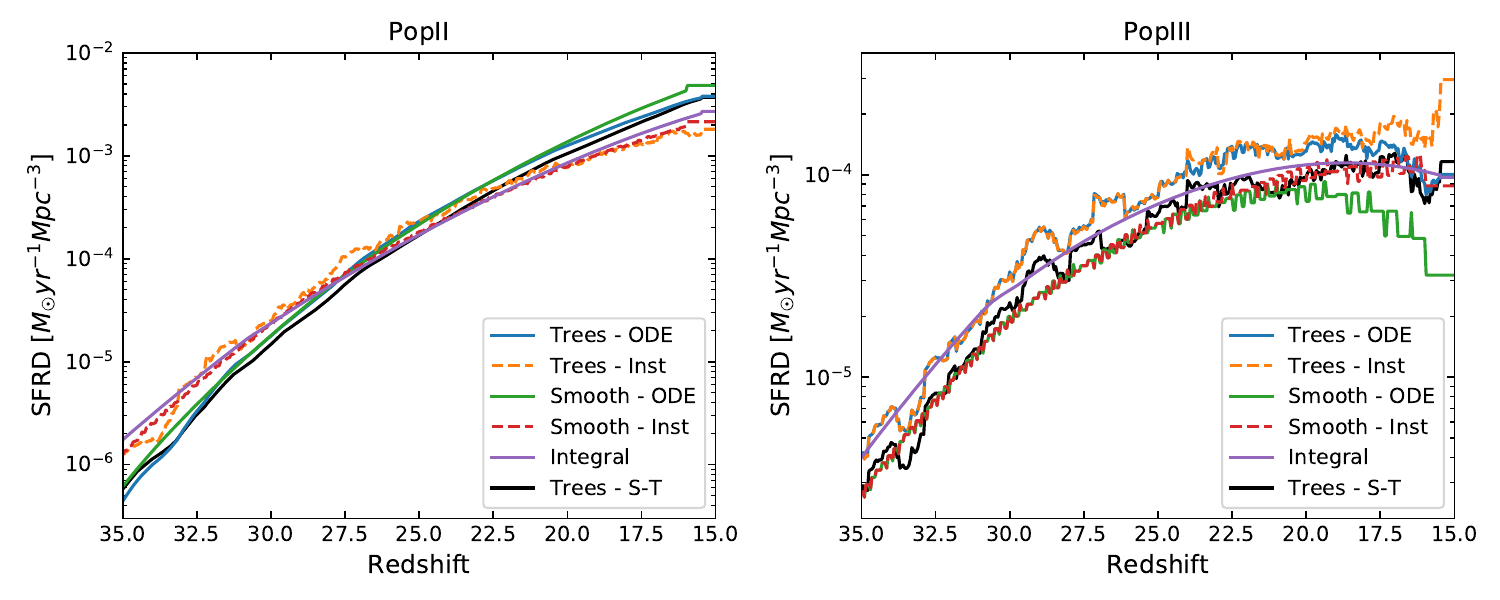}
    \caption{Time-evolution of the global PopII (left) and PopIII (right) SFRDs for different combinations of the methods used for determining halo merger history and for determining PopII star formation. Models utilizing the EPS merger tree method are denoted by “Trees” in the legend, and those utilizing the smooth accretion method for halo mass growth (equation \ref{EQ-SmoothAcc}) are labeled “Smooth”. Models that use the fiducial set of ODEs for PopII star formation (equations \ref{EQ-M*ODE} \& \ref{EQ-MgasODE}) are labeled as “ODE”, and those using the Instantaneous method (equation \ref{EQ-InstPopII}) are denoted by “Inst” in the legend. Also shown is the fully analytic Integral method (purple), as described in Section \ref{subsec:AltModels} with a PopII SFE of $f_{\rm II}=0.005$ and a new PopIII stellar mass of $M_{\rm III,new} = 400$ \Msun. Finally, the black ``Trees - S-T" curve uses the fiducial ODE PopII star formation method and EPS merger trees for the DM halo evolution, but the halo mass bin number densities, $n_{i}$, are weighted to match the Sheth-Tormen halo mass function at each redshift step, as is the case for the smooth accretion models. As discussed in the main text, we find that the smooth accretion gives similar results to the full merger tree models. Additionally, we find that instantaneous PopII SFRDs change less rapidly with redshift than our fiducial ODE-based prescription.}
    \label{fig:DM_SF_SFRDs}
\end{figure*}

Starting with our ``smooth accretion model'' of DM halo evolution,  we adopt the abundance matching model described in \cite{Furlanetto17}, 
\begin{equation} \label{EQ-SmoothAcc}
    \int_{m_{1}}^{\infty} n_{\rm h}(m|z_{1}) \,dm = \int_{m_{2}}^{\infty} n_{\rm h}(m|z_{2}) \,dm ,
\end{equation}
where $n_{\rm h}(m|z_{i})$ is the number density $n_{\rm h}$, of halos with masses $\geq m$, at redshift $z_{i}$, as determined by the Sheth-Tormen halo mass function \citep{ShethTormen}. For a given starting mass $m_{1}$ at redshift $z_{1}$, we determine the halo mass growth history for a constant $n_{\rm h}$ by solving for $m_{2}$ at each subsequent time step \citep[see][for a recent example]{Yamaguchi22}. 

Figure \ref{fig:SA_vs_MC_Halos} shows an example smooth accretion halo mass growth track along with a sampling of MC merger trees with the same final halo mass. While the smooth accretion model follows a constant cosmic number density through time for each halo mass, MC merger trees initialize $n_{\rm h}$ at $z = 15$ and determine the merger history for a given halo through EPS formalism. The two models produce halo mass evolution tracks that differ on average by a factor of $\sim$2 at $z \lesssim 35$ due to this discrepancy between Sheth-Tormen and EPS formalism. We note, however, that this discrepancy in the halo mass growth between models is outweighed by the uncertainties in other high redshift astrophysical parameters implemented here, and thus represents a viable comparison to our fiducial model. We intend to explore alternative DM halo evolution models in future work.

For the ``instantaneous" star formation prescription, we compare the SFRDs of our fiducial model with those determining enriched star formation via
\begin{equation} \label{EQ-InstPopII}
    M_{\rm *,II} = f_{\rm II, i} {M}_{\rm gas,new} .
\end{equation}
Here, ${M_{\rm *,II}}$ is the newly-formed PopII stellar mass, $f_{\rm II, i}$ is the instantaneous PopII SFE, and ${{M}_{\rm gas,new}}$ is the gas mass that has accreted into the halo over the last time step. This method uses a fraction, $f_{\rm II, i}$, of the gas mass accreting onto a halo at a given time step to instantly form PopII stellar mass within it. This means that if a halo already has PopII stars, it will not form any more PopII stellar mass until a halo with no prior enriched star formation merges with it and introduces new gas. In the case of two halos with previous PopII star formation merging, no new stellar mass is formed. We include this method into our comparison as it is implemented by semi-analytic models throughout the literature in various forms that resemble equation \ref{EQ-InstPopII}; using a SFE and gas mass to instantly introduce new stellar mass \citep[e.g.][]{Sun16, Park19, Magg18, Magg22}. For the rest of this paper, we refer to this PopII star formation prescription as the ``instantaneous" method.

For completeness, we also consider the analytic model detailed in \cite{Visbal15b} to compare with the results of our semi-analytic framework. This approach determines the SFRD by integrating the Sheth-Tormen halo mass function, $dn/dM$, at each time step \citep[also see][for another example of this method]{Mashian16, Munoz22, Munoz23}. Halo mass ranges are defined for both PopII and PopIII such that the cosmic mass fraction collapsed into DM halos, $F_{\rm coll}$, can be determined for stellar population $i$ via 
\begin{equation}
F_{\mathrm{coll},i}(z) = \frac{1}{\Omega_{\rm m}\rho_{\rm c}} \int_{M_{l,i}}^{M_{u,i}} M \frac{dn}{dM}(z) \ dM .     
\end{equation}
Here, $\Omega_{\rm m}$ is the cosmological density parameter for matter and $\rho_{\rm c}$ is the critical density. In this work, we  determine $F_{\rm coll}$ at each time step by integrating the halo mass function from \Mcrit--2.5\Mcrit \ for PopIII, and from 2.5\Mcrit--$10^{9}$\Msun \ for PopII. The time derivatives for both stellar populations, $(dF_{\rm coll}/{dt})_{i}$, are then calculated analytically at each step and are multiplied by $\rho_{\rm b}f_{\rm II}$ and $\Omega_{\rm m}\rho_{\rm c}{M_{\rm III,new}}$ to give the PopII and PopIII SFRDs, respectively. To more closely match the semi-analytic SFRD values, we double the fiducial value of both the PopII SFE and the new PopIII stellar mass in this analytic integration model, i.e. $f_{\rm II}$ = 0.005 and ${M_{\rm III,new}}$ = 400 \Msun. We also note that we smooth the value of $F_{\mathrm{coll}, i}$ over the previous ten time steps ($\Delta z = 0.5$) to avoid numerical feedback effects when determining its derivative.

\subsection{Comparison of Methods} \label{subsec:fig6}
Figure \ref{fig:DM_SF_SFRDs} shows the global SFRDs by stellar population for various combinations of DM halo model and star formation prescription, as well as the SFRDs of the fully analytic approach. A key takeaway of Figure \ref{fig:DM_SF_SFRDs} is that the merger trees and smooth accretion models for DM halo evolution give similar PopII SFRDs. For a given PopII star formation prescription, instantaneous or ODE, the PopII SFRDs of both DM halo models agree to within a factor of two over the redshifts shown.

We also find that the fiducial ODEs for enriched star formation yield steeper PopII SFRD growth than the instantaneous treatment. Calculating the total change in PopII SFRD over the change in time shown in Figure \ref{fig:DM_SF_SFRDs} (i.e. $\delta {\dot \rho}_{\rm *,II} / \delta t$), the slopes given by the ODE prescription are steeper than those of the instantaneous prescription by over a factor of two, given the same DM halo model. In our fiducial prescription, an isolated halo will form stars at a continually decreasing rate, whereas the instantaneous model requires fresh infalling gas, hence the shallower growth of the instantaneous PopII SFRDs.

Looking to the right panel of Figure \ref{fig:DM_SF_SFRDs}, we find that the PopIII SFRDs of the merger trees have similar qualitative behavior to the smooth accretion SFRDs, but are roughly double their value over the entire redshift range shown. This is a consequence of the discrepancy in halo number densities between Sheth-Tormen and EPS formalism as described in Section \ref{subsec:AltModels}, causing the smooth accretion models to underestimate the PopIII SFRDs by a factor of $\sim$2. We verified this by determining the halo mass function given by our EPS merger trees at each redshift then calculating its ratio with respect to the Sheth-Tormen halo mass function. We then used these ratio values to weight the SFRs of each individual halo before calculating the SFRD, effectively constraining the halo number densities of each mass bin to match the Sheth-Tormen halo mass function throughout the simulation. The resulting PopIII SFRD in Figure \ref{fig:DM_SF_SFRDs} (black) agrees with those of the smooth accretion model to within 15\% between $z = 20-35$. At lower redshifts the Smooth -- ODE curve decreases due to its higher PopII SFRD and LW background (discussed further in section \ref{subsec:HighLW}), and the Sheth-Tormen curve approaches the fiducial SFRD values as the weighting ratios approach unity at $z = 15$.

Looking to the SFRDs determined by analytic integration in Figure \ref{fig:DM_SF_SFRDs}, we see overall reasonable agreement with the SFRDs of the semi-analytic models. In fact, when compared to the fiducial SFRDs, the analytic values agree to within a factor of four and to within $\sim65\%$ over the redshifts shown for PopII and PopIII, respectively. The integral PopII SFRD shows better agreement with the instantaneous star formation models, falling within a factor of two and to within 40\% over the plotted redshifts for the merger trees and smooth accretion models, respectively. This agreement, however, is a result of tuning the star formation in the analytic integration model to match the results of our semi-analytic model. We used a PopII SFE and new PopIII stellar mass that were each twice their fiducial value, and set the boundary between PopIII and PopII star formation to be 2.5\Mcrit($z$). Variations of these parameter values can alter the resulting SFRDs by factors of a few. So while it is able to rapidly give order of magnitude agreement with the SFRDs of our semi-analytic framework, purely analytic models miss out on key astrophysics for early star formation, and so tuning is required to match semi-analytic values.
 
\begin{figure}
    \centering
    \includegraphics[width=0.5\textwidth]{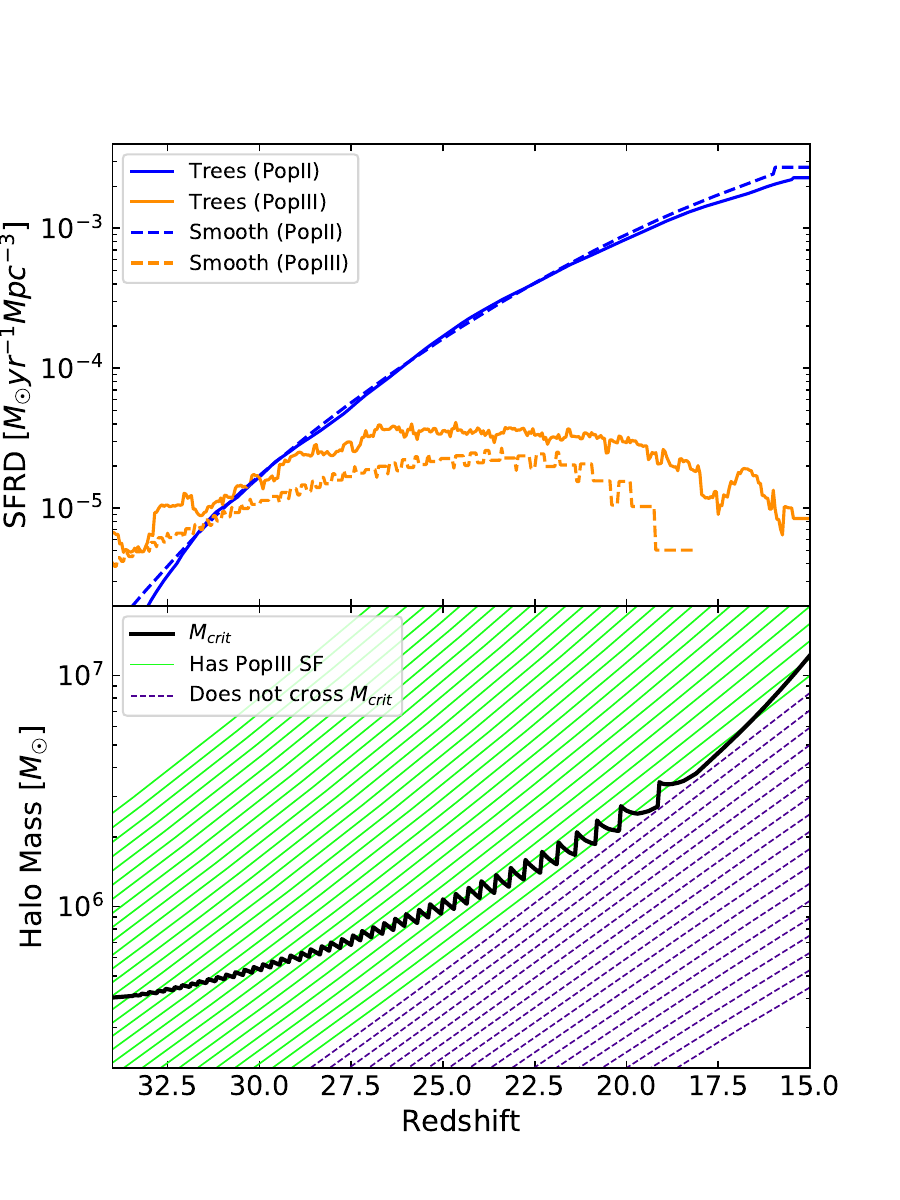}
    \caption{The breakdown of smooth accretion DM halo models in high LW background intensity cases. \textbf{Top:} The PopII (blue) and PopIII (orange) SFRDs for models utilizing MC merger trees (solid) and smooth accretion (dashed) for DM halo mass evolution, and an artificially boosted LW background of $10 J_{\rm LW}$. \textbf{Bottom:} The critical mass for the smooth accretion model shown in the top panel (thick black) alongside a sample of the halo mass growth tracks used in said model. The mass growth tracks that cross \Mcrit \ at any point (and therefore form PopIII stellar mass) are shown in green, and those that do not cross \Mcrit \ at all are shown by the purple dashed lines.}
    \label{fig:High_LW_SFRDs}
\end{figure}

\subsection{High $J_{\rm LW}$ Limiting Smooth Accretion Models}\label{subsec:HighLW}
In Figure \ref{fig:DM_SF_SFRDs}, we see a clear downturn in the ``Smooth -- ODE" PopIII SFRD at $z \lesssim$ 20 (green, right panel). Although this effect is also seen in the fiducial PopIII SFRD (blue, same panel), it is much more pronounced for the smooth accretion model than it is for the merger trees. This behavior results from a limitation of the smooth accretion model that we have discovered in cases of high LW background intensity. To illustrate the breakdown that smooth accretion models experience in high \JLW \ environments, we artificially increase the LW background by a factor of ten and compare the resulting SFRDs for merger trees and smooth accretion halos in Figure \ref{fig:High_LW_SFRDs}. 

The top panel of Figure \ref{fig:High_LW_SFRDs} shows the PopII and PopIII SFRDs for both MC merger trees and smooth accretion halo models, given a LW background intensity an order of magnitude higher than in our fiducial framework. The bottom panel shows the critical mass of the smooth accretion realization on top of a sample of the smooth accretion halo mass growth tracks used. The green mass tracks depict halos that cross \Mcrit \ at any single point in the simulation, whereas the dashed purple mass tracks do not ever cross \Mcrit. 

With a boosted LW background intensity, the critical mass increases more rapidly than in our fiducial model. At $z \sim$ 19, the last smooth accretion track crosses \Mcrit, after which no further PopIII star formation occurs. This suggests that smooth accretion models for DM halo mass evolution break down in high-radiation environments, and are unable to form stars as the growth of \Mcrit \ outpaces the growth of the halos. This is particularly relevant for \Mcrit \ models with a $z$-dependence that scales with \JLW, as in the \cite{Kulkarni21} model. MC merger trees, with their stochastic halo mass growth histories, cover a wider range of halo masses at all time steps, making the rapid growth of \Mcrit \ a nonissue for MC merger tree models. Thus, the complete termination of PopIII star formation in high LW environments represents a limitation to semi-analytic models with smooth halo mass growth determined via cosmic number density abundance matching.

\begin{figure*}
    \centering
    \includegraphics[width=\textwidth]{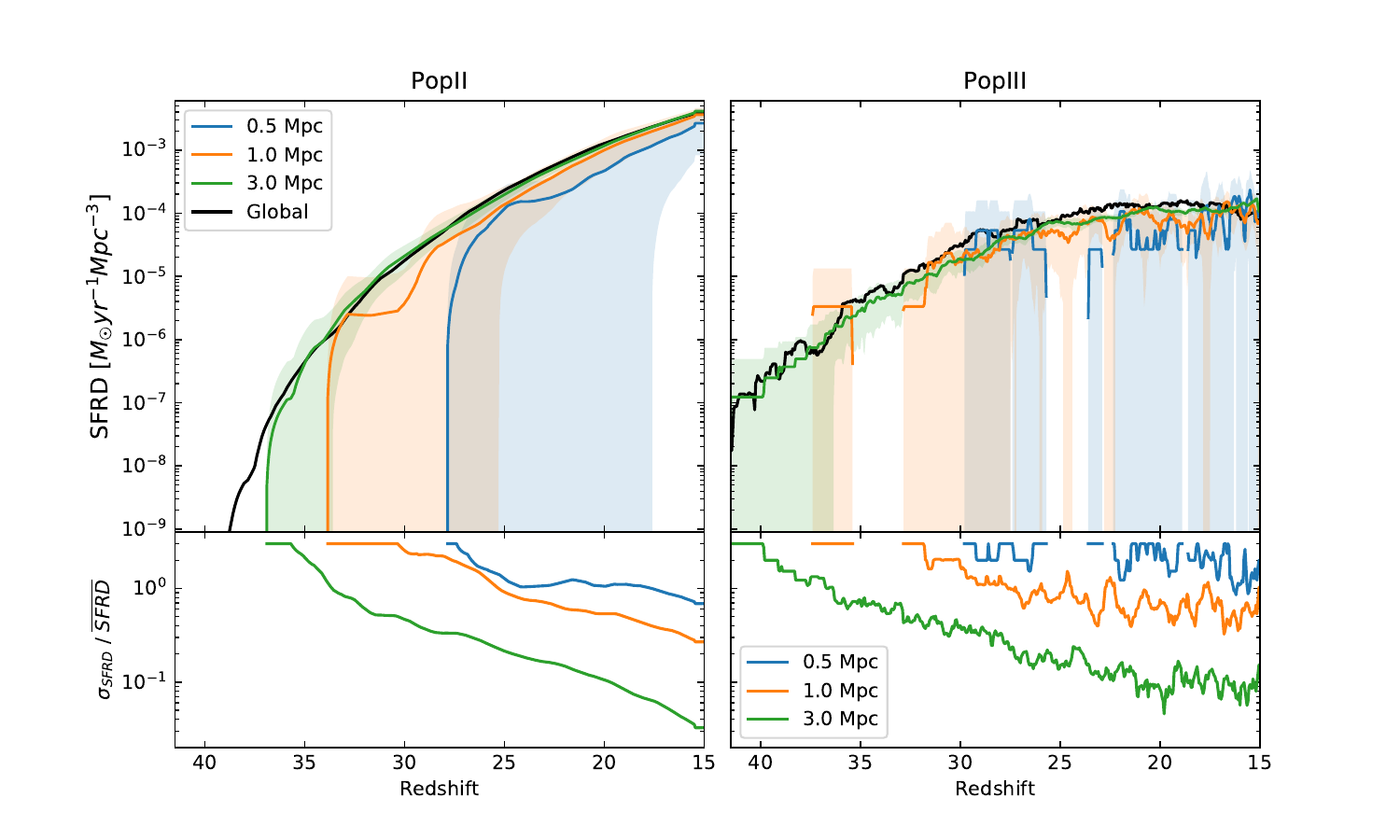}
    \caption{\textbf{Top panels:} The average PopII (left) and PopIII (right) SFRDs for a range of box sizes. Each curve is a 10-realization average and is denoted by the length of one side of the box volume. The shaded regions corresponding to each curve represent the standard deviation of the ten constituent runs. Also shown are the SFRDs of our fiducial model (black, both panels), which is a globally-averaged model and hence, a supposed upper limit for an ever-increasing box size. \textbf{Bottom panels:} The error on the mean for each of the PopII (left) and PopIII (right) SFRDs, i.e. the ratio of the standard deviation to the average for each corresponding curve in the top panels.}
    \label{fig:BoxSize}
\end{figure*}

\section{Results III: Realistic Box Volume Simulations} \label{sec:resultsIII}
In addition to studying the global star formation history, we also use our model to represent sub-volumes of the Universe. We accomplish this by building ``boxes" that simulate an ensemble of halos with realistic merger histories for a specified volume. While we cannot model spatial information within these boxes (e.g., the 3D positions of halos),  we can compute the variation from region-to-region due to differences in halo number and assembly history. As mentioned above, this is relevant to the large-scale modelling of spatial fluctuations in the high-redshift 21cm signal which are sourced by fluctuations in star formation. This approach also allows us to determine the extent to which finite simulation boxes (e.g., for hydrodynamical cosmological simulations) systematically underestimate the SFRD due to their lack of density fluctuations on spatial scales larger than the simulation box.

To simulate a finite-volume box, we begin by determining the mean number of halos in each mass bin (which are the same as described in Section \ref{sec:methods}) at the final redshift of our simulation, $z=15$. This number is computed using the Sheth-Tormen halo mass function with a power spectrum truncated (i.e. set to zero) at $k_{\rm min} = 2\pi/(\sqrt{3}L_{\rm box})$, corresponding to the longest length scale in the box. This truncation is applied to approximate the impact of boxes with density equal to the cosmic mean as is typically assumed for numerical cosmological simulations. We note that in future work we intend to explore simulating similar boxes with different mean overdensities, which could then inform the sub-grid modelling in large-scale semi-numerical models of the 21cm signal \citep{2012Natur.487...70V, Fialkov13, Magg22}. We sample the number of halos for each mass bin from a Poisson distribution and generate the MC merger trees for each as described in Section \ref{subsec:DMmethods}. Our fiducial semi-analytic model is then applied to these merger trees assuming an external LW background, $J_{\rm LW}(z)$, computed from our global fiducial model.

We present the results of our finite-volume boxes in Figure \ref{fig:BoxSize}. Here we show the mean PopII and PopIII SFRDs for 10 boxes of various volumes, and compare them to the SFRDs of our fiducial global model presented above. For boxes smaller than 3 Mpc across, we find that removing power associated with scales larger than the boxes results in a systematic reduction of the PopII SFRD. The PopIII SFRD however, is less impacted due to it being sourced by smaller halos (with abundance set by the power on smaller scales). From this, we conclude that a 3 Mpc box is sufficiently large to provide an unbiased estimate of the global star formation at $z=15-35$. One caveat is that this assumes an accurate model of $J_{\rm LW}$ (i.e. from our global fiducial semi-analytic simulation) and would not hold if the LW background were computed self-consistently from the finite-volume box. We also note that if the SFRD were to be dominated by more massive halos than in our model, larger-volume boxes would be required for an unbiased SFRD (as likely occurs at lower redshift).

In Figure \ref{fig:BoxSize}, we also show the scatter of the PopII and PopIII SFRDs for different box volumes. In order to prevent the scatter from depending on our redshift step size, we assign each PopIII star formation burst to a random time between $z$ and $z + \Delta z$, then smooth the results over a 3 Myr period (roughly corresponding to a typical PopIII stellar lifetime). This process effectively allows for star formation to proceed continuously over time instead of at discrete redshift values which bin star formation events into finite time spans, thereby introducing a dependence on the bin size (i.e. $\Delta z$) into the error calculations. As expected, the SFRDs increase with higher redshift as halos become increasingly rare. We also note that the fluctuations in PopIII are significantly higher, in large part due to PopIII star formation occurring in instantaneous bursts rather than smoothly across time as is the case for PopII in our model. The scatter is the error one would expect in a single simulation of the given box size. Additionally, these are the levels of scatter one expects in sub-regions of large-scale semi-numerical simulations, as these works typically assume $({\rm 3 \ Mpc})^3$ volume resolution elements corresponding to the scales where $v_{\rm bc}$ does not vary spatially. We find order unity scatter at $z\sim35$ and 10\% at $z\sim22.5$ for the PopII SFRD. For the PopIII, we predict order unity at $z\sim30$ and a value above 20\% for all our redshifts ($z>15$). 

In the future, we plan to extend this analysis to overdense/underdense regions to improve the modelling of large-scale 3D spatial fluctuations in star formation history. We note that the size fluctuations shown here represent lower limits because we assume fixed star formation efficiencies (i.e. constant $f_{\rm II}$ and $M_{\rm III, new}$). In reality these will vary from halo to halo, and vary over time for a specific halo. In future work, we intend to calibrate the SFEs with hydrodynamical cosmological simulations including star formation and radiative feedback (Hazlett et al., in prep).

\section{Star Formation History Comparison with Other Works} \label{sec:discussion}

\begin{figure*}
    \centering
    \includegraphics[width=\textwidth]{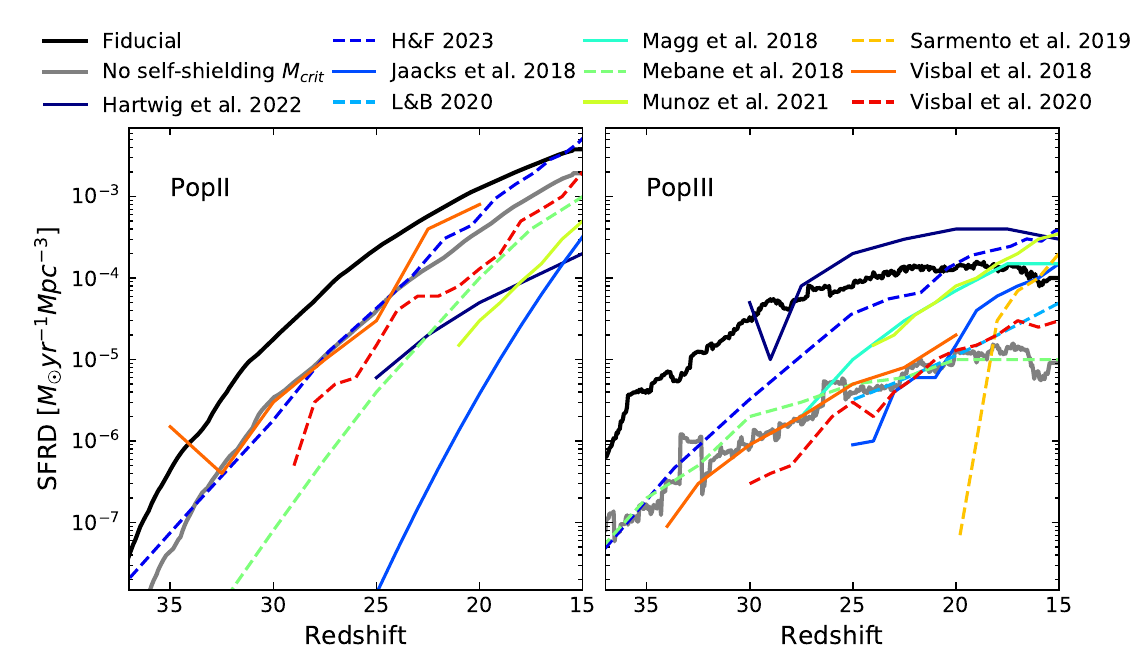}
    \caption{A comparison of the PopII (left) and PopIII (right) SFRDs found in this work with those from the literature. We show our fiducial SFRDs (thick black) and those found using the no self-shielding critical mass model (thick gray, see Figure \ref{fig:Mcrit-model}) alongside recently published SFRD histories. These SFRDs alternate between dashed and solid curves for clarity, and are labelled alphabetically by first-author surname. Note, for conciseness, \cite{Hegde23} and \cite{Liu20} are labelled as ``H\&F 2023" and ``L\&B 2020", respectively. We also note that not all referred works include PopII SFRD histories.}
    \label{fig:SFRD_Compare}
\end{figure*}

We now look to previous works throughout the literature and compare their SFRD histories to those found with our semi-analytic framework. Figure \ref{fig:SFRD_Compare} shows the PopII and PopIII SFRDs determined by our fiducial model, as well as those using the no self-shielding critical mass (see Figure \ref{fig:Mcrit-model}), compared to several recently reported SFRDs from the literature. These SFRDs are the results of a variety of modelling prescriptions including semi-analytic simulations \citep[e.g.][]{Magg18, Mebane18, Visbal18, Visbal20, Liu20}, cosmological hydrodynamic frameworks \citep{Jaacks18, Sarmento19}, and even a fully analytic model \citep{Munoz22}. We note that the majority of the curves shown in Figure \ref{fig:SFRD_Compare} are approximations taken from the fiducial SFRD evolution of each reference. This was done by gathering integer redshift SFRD values from plots in each published work, meaning that only broad trends are captured. Exceptions include \cite{Jaacks18} PopII SFRD which has an assumed functional form, \cite{Liu20} provide a fit for their PopIII SFRD, and the authors of \cite{Hegde23} kindly provided us with raw SFRD values resulting from their fiducial framework.

In the right panel of Figure \ref{fig:SFRD_Compare}, we see that the majority of the PopIII SFRDs shown fall below our fiducial SFRD; however, many of them come into agreement when compared with our no self-shielding \Mcrit \ realization. The SFRDs reported by \cite{Mebane18}, \cite{Visbal18}, \cite{Liu20}, and \cite{Visbal20} all agree with the no self-shielding \Mcrit \ PopIII SFRD to within a factor of a few throughout the redshift range shown. In particular, \cite{Mebane18} and \cite{Visbal18} are consistent to within a factor of $\leq$2 across the entire redshift range, despite their respective use of smooth accretion and N-body prescriptions for DM halo evolution. We therefore attribute the general agreement of these semi-analytic models with our no self-shielding \Mcrit \ PopIII SFRD to their use of the same prescription for the critical mass (equation \ref{EQ-Machacek}).

Focusing on a few particular cases in Figure \ref{fig:SFRD_Compare}, the PopIII SFRD found by \cite{Hartwig22} shows qualitative agreement with our fiducial model at $z\lesssim27$, albeit a factor of $\sim$3 higher. Although they adopt a critical mass model with \Htwo \ self-shielding \citep{Schauer21}, their \JLW \ is predetermined as a function of redshift which sets their \Mcrit \ to five times the threshold found in our self-consistent model at $z = 35$. Once PopIII star formation begins, however, we infer that their high SFE of $f_{\rm III} = 0.38$ per freefall time (calibrated from observations), paired with the limited negative feedback of their analytic \JLW($z$) prescription, gave this overall higher PopIII SFRD.

Most recently, the semi-analytic model presented in \cite{Hegde23} has one of the steepest PopIII SFRD evolutions shown in Figure \ref{fig:SFRD_Compare}. While their critical mass agrees with our fiducial \Mcrit \ to within a factor of a few throughout, their star formation prescription allows for multiple generations of PopIII stars per halo, a feature that most likely sources such rapid growth.

The PopIII SFRDs resulting from cosmological hydrodynamic simulations in Figure \ref{fig:SFRD_Compare} \citep[i.e.][]{Jaacks18, Sarmento19} are also significantly steeper than those of our semi-analytic framework. Since these models mainly cover lower redshifts than what is explored here (each ending at $z \sim $7), their SFRDs at the redshifts shown in Figure \ref{fig:SFRD_Compare} are likely products of the largest halos found in the simulation. Without resolving minihalos at higher redshifts, star formation is confined to the few resolved DM halos that persist at such early times, causing the SFRD to fall off more rapidly. If unresolved minihalos were included, they likely would supplement early star formation and carry the SFRDs at higher redshifts to give more qualitative agreement with our model.

Now looking to the left panel of Figure \ref{fig:SFRD_Compare}, we see the available corresponding PopII SFRDs which result from a diverse range of star formation prescriptions. On top of this diversity, PopII star formation is frequently affected by other modelling aspects as well (such as the \Mcrit \ model, $f_{\rm II}$, \tdelay, SNe feedback, etc.), which makes diagnosing variations between the resulting SFRDs more challenging than with PopIII. For example, \cite{Hegde23} determine PopII star formation using our fiducial ODEs adopted from \cite{Furlanetto21}, yet their PopII SFRD grows more consistently than the SFRDs resulting from our framework. This is likely a result of their PopII SFE which is determined at each time step using a star formation duty cycle, unlike the constant $f_{\rm II}$ value used here. Conversely, the SFRD of \cite{Munoz22} is determined via analytic integration of the halo mass function which, as discussed in Section \ref{subsec:fig6}, heavily depends on the halo mass ranges over which one integrates and the SFEs used. The PopII SFRDs given by \cite{Mebane18}, \cite{Visbal18, Visbal20}, and \cite{Hartwig22} each arise from prescriptions similar to our instantaneous PopII star formation model, but most likely differ in evolution as a result of varying PopII SFEs, varying the halo gas mass used, the inclusion of bursty star formation, and the DM halo model itself.

The SFRDs resulting from this research have extensively shown that one's modelling choices can significantly alter the resulting star formation history. In particular, the right panel of Figure \ref{fig:SFRD_Compare} tells us that one must use the improved \Mcrit \ from \cite{Kulkarni21} or similar works \citep[e.g.][]{Schauer21, Nebrin23} that accurately accounts for the effects of \Htwo \ self-shielding and baryon-DM streaming. In general, varying the SFEs of either stellar population can shift their SFRD values up or down with time, and the longer star formation is delayed, the steeper the SFRD growth once it commences.

\section{Conclusions \& Future Work} \label{sec:conclusions}
In this paper, we presented a new global semi-analytic model of the first stars and galaxies at redshifts $z \geq 15$. Our model includes complete DM halo merger histories (typically not included in global star formation models at such high redshifts) and is calibrated to the latest numerical simulations \citep{Kulkarni21}. Here we conclude by summarizing key results from our study and mention directions of future work. 

We found that, compared to previous calibrations in the literature, the updated critical halo mass significantly changes the high-redshift SFRDs (mainly due to lower \Mmol \ values as a consequence of the molecular hydrogen self-shielding). For instance in our fiducial model, the PopIII SFRD is increased by more than an order of magnitude compared to the no self-shielding critical masses based on the simulations of \cite{Machacek01, Greif11, Stacy12}. The PopII SFRD is increased by roughly one order of magnitude at $z=35$, but falls to a factor of two disagreement by $z=15$ (see Figure \ref{fig:Mcrit-model}).

We also assessed the impact of including the scatter in \Mmol \ due to individual differences in halo merger history and geometry between halos from the simulations of \cite{Kulkarni21}. Figure \ref{fig:Scatter} shows that including this scatter modestly increases the abundance of stars formed. It increased the final PopIII stellar mass density by a factor of $\sim$1.5, and the final PopII SMD only increased by $\sim$10\% relative to the no scatter case.

We also studied the effect that DM halo mergers have on star formation history by comparing the SFRDs resulting from MC merger trees to those resulting from smooth accretion based on abundance matching, as described in \cite{Furlanetto17} (Figure \ref{fig:SA_vs_MC_Halos}). For our fiducial astrophysical parameters, we found that the resulting SFRDs (both PopIII and PopII) each agreed to within a factor of two from $z$ = 18--31 for these DM halo models (see Figure \ref{fig:DM_SF_SFRDs}). This suggests that the merger history is likely not required to model global SFRDs, provided one does not need an accuracy much greater than a factor of two, but it is required for spatial fluctuations such as those we show in Figure \ref{fig:BoxSize}. We note, however, that our model does not incorporate changes in astrophysics due to mergers (e.g., an enhanced PopII star formation efficiency following mergers), which could occur in more sophisticated star formation models. We also point out that in models with very high \JLW \ (for instance if the star formation efficiency was significantly higher than we assume), smooth accretion models have an unphysical shutoff in primordial star formation and a detailed treatment of the merger history may be necessary (see Figure \ref{fig:High_LW_SFRDs}).

Additionally, we tested how the global star formation history is affected by our PopII star formation prescription. We compared our fiducial model, based on equations \ref{EQ-M*ODE} and \ref{EQ-MgasODE}  \citep{[from ][]Furlanetto21} to a simpler ``instantaneous'' method used previously in the literature and found that the instantaneous model leads to a shallower metal-enriched SFRD slope throughout the simulation (see Figure \ref{fig:DM_SF_SFRDs}). Our fiducial prescription yields steeper PopII SFRDs because it allows halos to continue forming stars in isolation, unlike the instantaneous model which requires infalling gas at each time step (resulting in relatively more late-time formation). The effects of this steeper enriched SFRD manifest in the PopIII SFRD at late times ($z \lesssim$ 25) when the increasing LW background intensity sufficiently increases \Mcrit \ to alter the PopIII star formation history as well. 

We also compared our fiducial merger-tree model to a simple integration of the halo mass function based on \cite{Visbal15a}. We found that this analytic calculation closely reproduces the PopII SFRD in the instantaneous prescription, and can reproduce our fiducial enriched SFRD to within a factor of $\sim$2. It does a more reasonable job for PopIII, agreeing with our fiducial PopIII SFRD to within a $\sim65\%$ at the redshifts we explored, but we again note that these SFRDs were the result of tuning the analytic prescription to more closely align with the semi-analytic SFRD values.

We showed a comparison of the SFRDs resulting from our semi-analytic framework to those from the literature in Figure \ref{fig:SFRD_Compare}, and found that one's modelling choices can significantly alter the resulting star formation history. In particular, we found that the reduction of \Mcrit \ due to the effects of \Htwo \ self-shielding and baryon-DM streaming allows for a PopIII SFRD that is about an order of magnitude higher than without the inclusion of such effects. We therefore recommend the use of the critical mass threshold from \cite{Kulkarni21} or similar works \citep[e.g.][]{Schauer21, Nebrin23} that accurately accounts for the effects of \Htwo \ self-shielding and baryon-DM streaming for predicting global SFRD evolution. 

However, since these works investigating the effects of \Htwo \ self-shielding are relatively recent, many uncertainties remain. Although \cite{Kulkarni21} and \cite{Schauer21} both utilize hydrodynamical cosmological simulations (\textsc{Enzo} and \textsc{Arepo}, respectively) to probe various redshifts, LW background intensities, and streaming velocities to determine the critical mass for star formation, the thresholds reported by the two works differ by a factor a few. While differences in the simulations used, the parameter spaces probed, and the assumed thresholds for halo cooling/collapse may have contributed to this, exact sources of this discrepancy remain unclear. More recently, \cite{Nebrin23} published their critical mass model that builds up from halo gas physics to include dependencies on \JLW \ and \vbc \ using (semi-)analytic methods. However, despite hopes that a more analytic approach could shed light on the discrepancy, their resulting \Mcrit \ seemingly falls between the two models of \citep{Kulkarni21} and \cite{Schauer21}. We therefore caution the reader to take care when selecting a particular critical mass threshold for future star formation models.

Finally, we used our model to simulate finite-volume boxes each with their own unique set of halos and merger histories (though not including 3D spatial positions). For boxes which are forced to have a density equal to the cosmic mean (as is typically done in non-zoom cosmological numerical simulations), we tested how the lack of power on scales larger the simulation box can impact estimates of the star formation history. We found that a $(3 \ \rm Mpc)^{3}$ box is sufficiently large to provide an unbiased estimate of the global star formation at $z=15-35$, provided that a reliable external LW background is known (for analytic fits of various LW backgrounds found by our semi-analytic model, see Appendix \ref{appendixLW}). We also determined the typical scatter in the PopIII/PopII SFRDs as a function of redshift in regions of various volumes that are set to the mean cosmic density.

In future work, we will apply our model to predict large-scale spatial fluctuations in high-redshift star formation, which can then be used to model the cosmological 21cm signal. We will extend our finite-volume box analysis presented in Section \ref{sec:resultsIII} to include variations in star formation as a function of local overdensity and baryon-dark matter streaming velocity in $(3~{\rm Mpc})^3$ volume elements \citep[as is typically done in other 21cm predictions, e.g.][]{2012Natur.487...70V, Fialkov13}. We will then explore ways to rapidly emulate our model, such that it is possible to self-consistently determine the LW background and associated feedback across ${\sim}$Gpc distance scales while also accounting for variations in halo abundance and merger history down to the critical halo mass for forming PopIII stars.

\begin{acknowledgments}
We acknowledge support from NSF grant AST-2009309 and NASA ATP grant 80NSSC22K0629. All numerical calculations were carried out at the Ohio Supercomputer Center (OSC). We thank Sahil Hegde for kindly providing us with SFRD data from \cite{Hegde23} for Figure \ref{fig:SFRD_Compare}.
\end{acknowledgments}

\bibliography{Bibliography}{}
\bibliographystyle{aasjournal}

\begin{appendix}

\section{Fits for the Global Lyman-Werner Background Intensity} \label{appendixLW}

\begin{figure*}
    \centering
    \includegraphics[width=\textwidth]{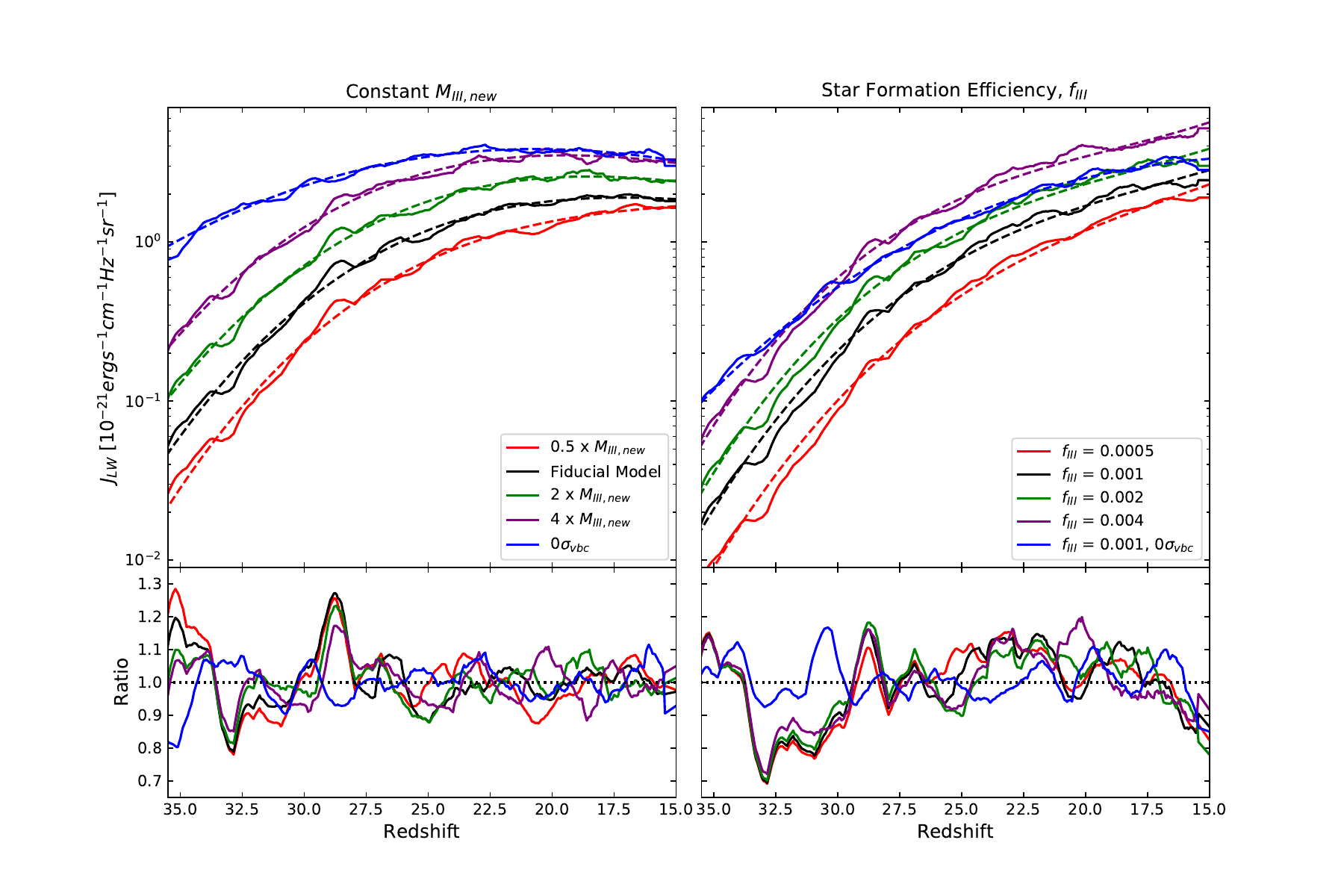}
    
    \caption{\label{fig:LW_fits} \textbf{Top Left:} Global Lyman-Werner background intensities for various realizations of our semi-analytic framework (solid) and their corresponding analytic fits (dashed). We show the \JLW($z$) values for our fiducial model (black) alongside those with 0.5$M_{\rm III,new}$ = 100 \Msun \ (red), 2$M_{\rm III,new}$ = 400 \Msun \ (green), 4$M_{\rm III,new}$ = 800 \Msun \ (purple), and for the case with no baryon-DM streaming velocity (blue). \textbf{Top Right:} Same as top left, but for realizations using a PopIII star formation efficiency as described in Section \ref{subsec:varyParams} (also see Figure \ref{fig:Parameters}). \textbf{Bottom Panels:} The ratios of each \JLW($z$) in the upper panels to the values given by their corresponding fits.}
\end{figure*}

Here we detail a sampling of analytic polynomial fits for different LW background intensities resulting from various realizations of our semi-analytic model. The fits provided here may be used in finite-volume models to estimate global star formation in the redshift range $z$ = 15--35. In the top left panel of Figure \ref{fig:LW_fits}, we show the \JLW($z$) resulting from realizations with various $M_{\rm III,new}$ values, and one for the case of no global baryon-DM streaming velocity (solid curves). We show the corresponding fit to each LW background by the identically colored dashed lines. Similarly, the top right panel shows the \JLW($z$) for realizations of our semi-analytic framework with PopIII star formation determined by a star formation efficiency (as described in Section \ref{subsec:varyParams}). The ratio of the model data to the fit of each \JLW($z$) is then shown in the bottom panels of Figure \ref{fig:LW_fits}, and we present the coefficients for each polynomial fit in Table \ref{tableOcoeffs}. For each realization, the coefficients A, B, C, and D describe the LW background intensity in terms of redshift $z$, via the following function:
\begin{equation} \label{EQ:LWfits}
    \log_{10}(J_{\rm LW}(z)) = A(1 + z)^{3} + B(1 + z)^{2} + C(1 + z) + D
\end{equation}

\begin{table*}
    \begin{tabular}{c c c c c}
        \hline
        Realization & A & B & C & D\\
        \hline\hline
        $0.5 \times M_{\rm III,new}$ & $-1.6246 \times 10^{-4}$ & $7.2219 \times 10^{-3}$ & $-0.11836$ & $0.93553$\\
        Fiducial Model & $-1.3808 \times 10^{-4}$ & $5.2787 \times 10^{-3}$ & $-5.5111 \times 10^{-2}$ & $0.36564$\\
        $2 \times M_{\rm III,new}$ & $-1.0885 \times 10^{-4}$ & $3.3920 \times 10^{-3}$ & $-8.1113 \times 10^{-3}$ & $8.2743 \times 10^{-2}$\\
        $4 \times M_{\rm III,new}$ & $-1.1377 \times 10^{-4}$ & $4.1009 \times 10^{-3}$ & $-2.4789 \times 10^{-2}$ & $0.30972$\\
        \vbc \ = 0 km $s^{-1}$ & $-2.2360 \times 10^{-5}$ & $-9.8214 \times 10^{-4}$ & $7.3987 \times 10^{-2}$ & $-0.33048$\\
        $f_{III}$ = 0.0005 & $-1.7990 \times 10^{-4}$ & $8.9799 \times 10^{-3}$ & $-0.20440$ & $2.0708$\\
        $f_{III}$ = 0.001 & $-2.0501 \times 10^{-4}$ & $1.0871 \times 10^{-2}$ & $-0.23545$ & $2.2763$\\
        $f_{III}$ = 0.002 & $-2.2604 \times 10^{-4}$ & $1.2679 \times 10^{-2}$ & $-0.28003$ & $2.7469$\\
        $f_{III}$ = 0.004 & $-2.5241 \times 10^{-4}$ & $1.4942 \times 10^{-2}$ & $-0.33499$ & $3.3205$\\
        $f_{III}$ = 0.001, \vbc \ = 0 km $s^{-1}$ & $-5.0559 \times 10^{-5}$ & $4.5826 \times 10^{-4}$ & $1.1234 \times 10^{-2}$ & $0.43310$\\
        \hline
    \end{tabular}
    \caption{Polynomial coefficients used in equation \ref{EQ:LWfits} for the \JLW($z$) fits shown in Figure \ref{fig:LW_fits}.}
    \label{tableOcoeffs}
\end{table*}
\end{appendix}

\end{document}